\documentclass[usenatbib]{mnras}
\newcommand{\pa}{\,\rlap{\raise 0.5ex\hbox{$\propto$}}{\lower 1.0ex\hbox{$\sim$}}\,}
\usepackage[dvips]{graphicx}
\usepackage{float}
\usepackage{epsf,graphics,graphicx}
\usepackage{rotating}
\usepackage{lscape}
\usepackage{amsmath}
\usepackage{hyperref}

\def\lsim{\!\!\!\phantom{\le}\smash{\buildrel{}\over
 {\lower2.5dd\hbox{$\buildrel{\lower2dd\hbox{$\displaystyle<$}}\over
                                 \sim$}}}\,\,}
\def\gsim{\!\!\!\phantom{\ge}\smash{\buildrel{}\over
{\lower2.5dd\hbox{$\buildrel{\lower2dd\hbox{$\displaystyle>$}}\over
                               \sim$}}}\,\,}
\begin{document}

 \title[VLA 5.5\,GHz observations of the GOODS-North Field] {The eMERGE Survey I: Very Large Array 5.5\,GHz observations of the GOODS-North Field}
 \author[D. Guidetti et al.]{D. Guidetti\thanks{E-mail: d.guidetti@ira.inaf.it}$^1$, M. Bondi$^1$, I. Prandoni$^1$,
T.W.B. Muxlow$^2$, R. Beswick$^2$,
\newauthor
N. Wrigley$^2$, I. Smail$^3$, I. McHardy$^4$
A.P. Thomson$^3$, J. Radcliffe$^{2,5,6}$, M.K. Argo$^{7,2}$
   \\
   $^1$ INAF - Istituto di Radioastronomia, via Gobetti 101, I--40129 Bologna, Italy\\
   $^2$ Jodrell Bank Centre for Astrophysics, Alan Turing Building, School of Physics and Astronomy, The University of Manchester, \\
        Oxford Road, Manchester, M13 9PL, U.K\\
   $^3$ Centre for Extragalactic Astronomy, Department of Physics, Durham University, South Road,
   Durham DH1 3LE, UK \\
   $^4$ Physics and Astronomy, University of Southampton, Southampton SO17 1BJ, UK \\
   $^5$ Kapteyn Astronomical Institute, University of Groningen, 9747 AD Groningen, The Netherlands \\
   $^6$ ASTRON, the Netherlands Institute for Radio Astronomy, Postbus 2, 7990 AA, Dwingeloo, The Netherlands \\
   $^7$ Jeremiah Horrocks Institute, University of Central Lancashire, Preston PR1 2HE, UK }

\date{Received }
\maketitle

\begin{abstract}

We present new observations of the GOODS-N field obtained at 5.5\, GHz with the
Karl G. Jansky Very Large Array  (VLA). The central region of the field
was imaged to a median r.m.s. of 3\,$\mu$Jy beam$^{-1}$ with
a resolution of 0.5\,arcsec.
From a 14-arcmin diameter region we extracted a sample of 94 radio sources
with signal-to-noise ratio greater than 5.
Near-IR identifications are available for about $\sim$88 percent of the radio sources.
We used different multi-band diagnostics to separate active galactic nuclei (AGN), both radiatively
efficient and inefficient, from star-forming galaxies.
From our analysis, we find that about 80 percent of our radio-selected sample
is AGN-dominated, with the fraction raising to 92 percent when considering only the
radio sources with redshift $>1.5$.
This large fraction of AGN-dominated radio sources at very low flux densities
(the median flux density at 5.5\,GHz is
42\,$\mu$Jy), where star-forming galaxies are expected to dominate,
is somewhat surprising and at odds with other results.
Our interpretation is that both the frequency and angular resolution of our radio observations
strongly select against radio sources whose brightness distribution is diffuse on scale of several kpc.
Indeed, we find that the median angular sizes of the AGN-dominated sources is around 0.2-0.3\,arcsec 
against 0.8\,arcsec for star-forming galaxies. This highlights 
the key role that high frequency radio observations can play
in pinpointing AGN-driven radio emission at $\mu$Jy levels.
This work is part of the eMERGE legacy project. 

\end{abstract}


\section{INTRODUCTION}
\label{sect-intro}

A complete census of both star-formation and nuclear activity over cosmic time is crucial to understanding 
the assembly and evolution of galaxies and super-massive black holes (SMBHs), as well as the role of mergers 
and secular processes in driving their growth. The similar evolution found for the comoving star-formation rate (SFR)
density and the comoving SMBH accretion rate, both peaking at z$\sim$2 \citep[e.g.][for a review] {2014ARA&A..52..415M}, 
and the tight correlations between galaxy properties and BH mass \citep[e.g.][]{1998AJ....115.2285M, 2000ApJ...539L..13G, 
2002ApJ...578...90F, 2009ApJ...706..404G} suggest a synchronized evolution likely driven by related physical processes 
\citep[e.g.][]{2005Natur.433..604D, 2007ApJ...659..976H, 2014MNRAS.441.1059V}. A key question is the role of Active 
Galactic Nuclei (AGN) in such scenarios, with AGN outflows possibly being responsible for regulating or terminating the 
star-formation \citep[][and references therein]{2014ARA&A..52..589H}.

Attempts to derive the star-formation and accretion histories through optical, near-IR (NIR) and X-ray surveys suffer 
significant uncertainties because of the large and mostly unconstrained corrections for dust extinction and gas obscuration.
Even the deepest X-ray surveys may fail to detect the most heavily absorbed Compton thick AGN. 
Many {\it Spitzer Space Telescope} and {\it Herschel Space Observatory} observations have been dedicated to detect the dust 
emission in galaxies up to z$\sim$2  \citep[e.g.][]{2013A&A...549A..59D}. However, these studies are affected 
by the poor angular resolution (from few arcsec to $\sim$ 30 arcsec) of IR telescopes which cannot resolve compact 
structures at high redshift and are confusion limited.
In contrast, radio continuum imaging is a powerful dust and obscuration-free tool providing
unbiased measures of both star-formation and AGN activity up to high redshift, and, moreover, interferometry 
techniques can reach sub-arcsec angular resolution, up to milli-arcsec scales through Very Long 
Baseline Interferometry (VLBI).

Increasing observational evidence suggests that the sub-mJy radio source population
is a mixture of star-forming galaxies (SFGs), radiatively efficient AGN (RE-AGN) and radiatively inefficient 
AGN (RI-AGN), with the formers dominating at the lowest flux densities below $\sim$100\,$\mu$Jy 
\citep[e.g.][]{1999MNRAS.304..199G, 2005MNRAS.358.1159M, 2006MNRAS.372..741S, 2007A&A...463..519B,
2008MNRAS.386.1695S, 2009ApJ...690..610S, 2013MNRAS.436.3759B, 2017arXiv170309719S}. 
The unexpected detection in deep radio surveys of large numbers of 
RE-AGN  \citep[e.g][]{2014ARA&A..52..589H}, emitting over a wide range of the electromagnetic spectrum, 
from mid-IR (MIR) to X-rays, but typically radio-quiet, has opened the exciting prospect of studying 
the whole AGN population through deep radio surveys.
The nature of the radio emission in RE-AGN is currently hotly debated. Several works suggest that 
star-formation related processes can, at least partly, produce the observed radio emission in RE-AGN 
\citep[e.g.][]{2011ApJ...739L..29K, 2011ApJ...740...20P, 2012ApJ...758...23C, 2013MNRAS.436.3759B}. 
Others point towards composite star-formation and AGN  emission \citep[e.g.][]{2015MNRAS.448.2665W}.
The  presence of embedded AGN cores has been demonstrated in some systems through VLBI observations 
\citep{2016A&A...589L...2H, 2016A&A...589L...3M}. 

Assessing the faint AGN component in deep radio fields will provide an important tool to understand the 
role of nuclear activity in distant galaxies, the nature and accretion regime of RI- and RE-AGN, and their 
possible co-evolution with star-formation processes. The most direct way to identify faint AGN-driven radio 
emission is the detection of embedded radio cores in the host galaxies, through
ultra-deep and high resolution radio observations, supported by multi-wavelength observations, crucial 
to understand the physical properties and nature of the radio sources and their hosts.

This context motivates the eMERGE survey \citep[e-MERlin Galaxy Evolution survey][]{2008evn..confE..22M}, 
the largest e-MERLIN (enhanced Multi-Element Remote-Linked Interferometer Network) legacy project, whose goal 
is to obtain a resolved view of the radio source population up to high redshift in the Great Observatories 
Origins Deep Surveys-North \citep[GOODS-N; ][]{2004ApJ...600L..93G}.
GOODS-N was observed previously at 1.4\, GHz \citep{2000ApJ...533..611R, 2005MNRAS.358.1159M, 2010ApJS..188..178M} 
and 8.5\,GHz \citep{1998AJ....116.1039R} with the Karl G. Jansky Very Large Array (VLA).
Preliminary observations at 5.5\,GHz were obtained as part of the e-MERLIN commissioning \citep{2013MNRAS.432.2798G}.
eMERGE is based on the combination of ultra-deep e-MERLIN and VLA  observations at 1.4 and 
5.5\,GHz. When completed, it will provide sub-$\mu$Jy sensitivity on 0.05-2\,arcsec angular scales, corresponding 
to sub-kpc up to tens of kpc linear scales at redshift $z>1$, and will allow the separation of compact AGN-related 
emission from more extended, lower surface brightness star-forming disks.

In this paper, we present the first  5.5\,GHz deep image and catalogue of the GOODS-N field based on VLA 
observations with sub-arcsec resolution, taken as part of the eMERGE legacy project. This first set of observations 
are used to make an exploratory analysis of the $\mu$Jy radio source population, as observed at sub-arcsec resolution, 
with a particular focus on the AGN population. Near-IR identifications were obtained from available ultra-deep 
$K_s$-band catalogues \citep{2010ApJS..187..251W, 2011PASJ...63S.379K, 2014ApJS..214...24S}
and different diagnostics were used to separate different classes of AGN from SFGs. 
A preliminary report of this work has been presented by \citet{2015fers.confE..23G}. 

In a forthcoming paper (hereafter referred to as Paper II) we will extend this analysis to the radio spectral 
index properties of a larger sample of sources selected at 1.4\,GHz, that will be used, in combination with 
the wealth of broad-band information available in the GOODS-N field, to further characterize the properties 
of different types of AGN and the population of SFGs.

\begin{figure}
\centering
\includegraphics[width=8.5cm]{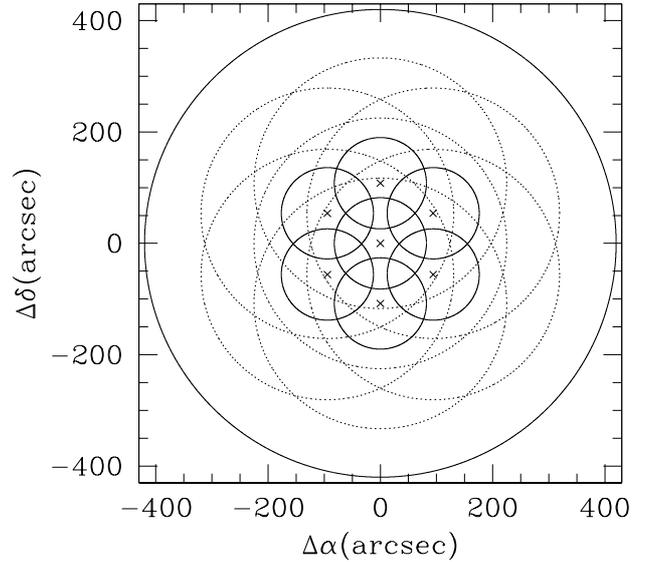}
\caption[]{Mosaic pattern of the 7 pointing centres (crosses) at 5.5\,GHz.
The dotted and inner full circles show the full width half power primary beam of the VLA ($\simeq$7.5\,arcmin)
and of the Lovell telescope ($\sim$2.5\,arcmin), respectively,
at this frequency. The outer full circle contains the area covered by our 5.5\,GHz catalogue.
The plot clearly illustrates that the region observed by the VLA alone
is oversampled to ease future combination with e-MERLIN observations
including the Lovell telescope.
}
\label{pattern}
\end{figure}

The paper is organised as follows. \S 2 describes the observations and the data reduction, while we present 
the catalogue extraction in \S 3. In \S 4 we provide the 
results on the polarisation analysis. NIR identifications and redshift information of the
radio sources are presented in \S 5. In \S 6 we identify AGN in our radio-selected sample using different diagnostics: 
various IR colour-colour plots, X-ray luminosity, radio-excess and VLBI detection. A discussion of the results is
presented in \S 7, while conclusions and future perspectives are given in \S 8.

Throughout this paper we adopt a concordance cosmology with Hubble constant $H_{0} =70\rm{km s^{-1}/Mpc^{-1}}$, 
$\Omega_{\Lambda} = 0.7$ and $\Omega_{M}= 0.3$. All magnitudes referred in this paper are in AB system, unless 
otherwise stated, where an AB magnitude is defined as AB $=23.9 - 2.5\log(S{\rm [\mu Jy]})$.

\section{VLA OBSERVATIONS AND DATA REDUCTION}
\label{sect-obs}

\subsection{Observations}

We obtained new VLA observations  of the GOODS-N field at a central frequency of 5.5\,GHz with a 2\,GHz 
bandwidth in A- and B-configurations. The VLA A-configuration observations were taken over four nights in October 2012 
(2012 October 6, 7, 8 and 20, project code 12B-181), for a total observing time of 14\,hrs. The B-configuration data 
were taken in one night (2013 September 27, project code 13B-152) for a total time of 2.5\,hrs.

The observations consist of a mosaic of 7 pointings (Fig.\,\ref{pattern}),
centred at $\alpha = 12^{\rm h}36^{\rm m}49^{\rm s}.4$,
$\delta = +62^{\circ}12^{\prime}58^{\prime\prime}$ (J2000).
The pointing centres are separated by $\sim $2\,arcmin to facilitate
combination with future 5.5\,GHz e-MERLIN observations including the 76-m Lovell telescope, 
which has a smaller primary beam (2.5\,arcmin) than the VLA antennas (7.5\,arcmin FWHM at 5.5\,GHz, see Fig.\,\ref{pattern}). 
For the VLA alone this mosaic pattern is oversampled, and provides ultra-deep sensitivity over the central region.
Each pointing was observed for a total of $\sim$80\,min in A-configuration and 12\,min in 
B-configuration. 
The nearby unresolved phase calibrator J124129+6020 was monitored for 40\,seconds every 10\,minutes to provide 
accurate phase and amplitude calibration. The flux density and bandpass calibrators, 3C\,286 and J1407+28279 
(OQ208), were observed once per night.

The data were recorded every 1\,s in spectral line mode using 16 adjacent $64\times2$\,MHz intermediate frequency 
channels (IFs), for a total bandwidth of 2048\,MHz. Both circular polarisations were recorded. 
This bandwidth synthesis mode minimises chromatic aberration (bandwidth smearing) and reduces the 
effects of narrow-band radio frequency interference (RFI) as individual narrow channels can be flagged and rejected 
from the data.

\subsection{Editing \& Calibration}
\label{sect-cal}

The data were calibrated and edited using the  {\sc aips} software package,
developed by the National Radio Astronomy Observatory\footnote{The National Radio
Astronomy Observatory is a facility of the National Science Foundation operated under cooperative agreement by 
Associated Universities, Inc.}, following standard procedures (as briefly outlined below). 
The five observing sessions (four in A-configuration and one in B-configuration) were edited 
and calibrated separately.

To set the interferometer group delay correctly we used the task {\sc fring} on the calibrator J1407+28297, 
selecting a time range of 1\,min. After a first run of automatic flagging done through {\sc rflag}, we 
performed a first calibration of the bandpass and of the flux density scale using the source 3C286. 
Amplitude calibration was based on the VLA standard spectrum of 3C286, bootstrapped to determine the spectrum 
of J1407+28279 and J124129+6020. Specifically, the frequent observations of J124129+6020 (once every 10 minutes) 
were used to calibrate the amplitudes and phases of the target fields. We then examined the visibilities of all 
the calibrators and performed further editing of residual RFI using the tasks {\sc spflg} and {\sc uvflg}.
A new calibration table was then derived using only the data that had passed the editing process. 
In total, about 15 percent of the {\it uv}-data were discarded in the editing process.

We obtained a mean flux density (averaged over the five days) of
2.271$\pm$0.032\,Jy for J1407+28279 and 0.298$\pm$0.003\,Jy for J124129+60200 (where the uncertainty is the standard 
error $\sigma/\sqrt{N}$), at the central observing frequency of 5.5\,GHz. We estimate calibration errors of 
$\sim$1-1.5 percent for the flux density measurements.

We also performed the polarisation calibration using the sources 3C286 and J1407+28279 (the latter is known to be 
unpolarised) as polarisation angle and instrumental polarisation calibrators, respectively. J1407+28279 was then imaged
in total intensity and polarisation to check for residual instrumental
polarisation. We derived a fractional polarisation, averaged over the five days,
of 0.2 percent, that we take as the level of residual instrumental polarisation
in our data.

Finally, the edited and calibrated {\it uv}-data from all five
observing sessions were combined for imaging using the task {\sc dbcon} with parameter {\tt REWEIGHT=1}.

\subsection{Imaging and mosaicing}
\label{sect-imaging}

Imaging wide fields using data sets with a large fractional bandwidth
($\Delta\nu/\nu_c$) is a challenging task given that the field of view, the primary beam correction, 
the synthesized beam or point-spread function, and the flux densities of the sources all vary significantly with frequency.

Two different approaches can be followed:
\begin{itemize}
\item{} Split the {\it uv}-data into sub-bands having
$\Delta\nu/\nu_c \ll 1$ (namely the sixteen IFs, each with a bandwidth of 128 MHz), and image separately 
each sub-band with a common resolution (tapering and/or changing the weight function with increasing frequency). 
At each frequency,  the mosaic, resulting from the combination of the seven pointing, is derived applying 
the primary beam correction appropriate to the central frequency of each sub-band. Finally,  the mosaics 
produced from each sub-band can be recombined with appropriate weights to obtain a sensitive wide-band 
mosaic \citep[e.g.][]{2012ApJ...758...23C}.

\item{} Imaging the entire bandwidth simultaneously using the multi-scale multi-frequency (MSMF) synthesis 
clean algorithm (available in the {\sc casa} package) with nterm $>1$, which takes into account frequency-dependent 
variations over the observing band \citep[e.g.][]{2011A&A...532A..71R, 2014arXiv1403.5242R}.
The resulting images for each pointing are then corrected for the primary beam using the {\sc casa} task 
{\sc widebandpbcor}. Finally, the mosaic covering the whole field is obtained by a weighted combination of the seven pointings.
\end{itemize}

We have tested and compared both approaches, and selected the MSMF synthesis clean for the following reasons. 
While both methods produce highly comparable images in terms of noise properties and image fidelity, the first method 
has, in our opinion, two main disadvantages. Firstly, the images produced by the different IFs have to be restored 
to the lowest common resolution. This means that we lose resolution in our images, but also that we need to 
fine-tune the data and/or change the weighting function used in the gridding process as the image
frequency increases. Secondly, the cleaning threshold is usually set to some
multiple of the expected noise (usually in the range 3 to 5). Using the same
criterion for the cleaning in both methods described above means that the
individual IF images will have, on average, a noise four times larger than the
image produced with the MSMF clean (since 16 individual IFs are summed up).
Therefore a large fraction of the sources detected in the sensitive wide-band
mosaic obtained by recombining the 16 individual IF mosaics will be sources that were not cleaned, or 
that were cleaned in some images (some IFs) and not in
others. This may affect the source properties by introducing subtle undesirable effects on the final mosaic.

We imported the combined data-sets (one for each pointing, including all the
A- plus B-configuration data) into {\sc casa} (task {\sc importuvfits}) and ran the
task {\sc clean} with two Taylor terms in the frequency expansion ({\tt mode=mfs, nterms=2}). Wide-field mode was enabled 
using {\tt gridmode=widefield, wprojplanes=128, facets=1}, along with three resolution scales. 
For each pointing a map with $8192$ pixels on-the-side was produced with a pixel 
size of $0.1\times 0.1$ arcsec$^2$.
The fields were cleaned down to about 4 times the expected r.m.s. noise of
each pointing. The final restoring beam was set to $0.56\times 0.47$ arcsec$^2$
with a position angle of $88^\circ$. After the deconvolution, wide-band primary beam correction was applied 
using the {\sc casa} task {\sc widebandpbcor} using a primary beam threshold cut-off of 0.15 per cent of the peak.

\begin{figure}
\centering
\includegraphics[width=8.5cm]{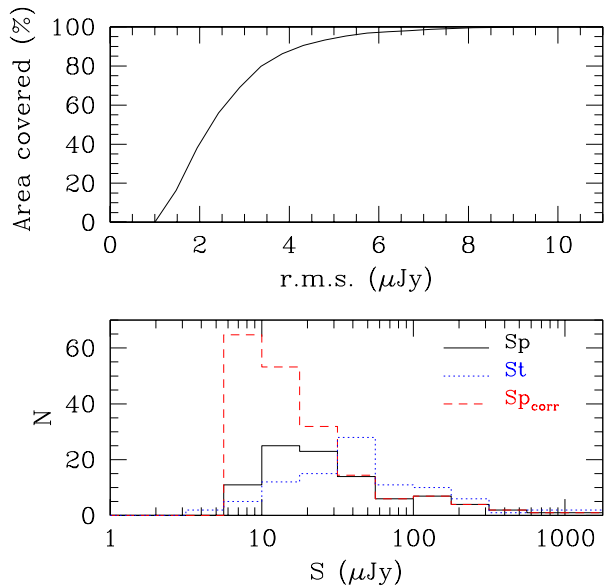}
\caption[]{Top: Visibility function of the 5.5\,GHz mosaic (area versus r.m.s. sensitivity) calculated over the region used 
to derive the source catalogue (see text). Bottom: Observed peak (S$_{\rm p}$) and total flux density (S$_{\rm t}$) 
distributions of the 5.5\,GHz sources. 
The dashed red line indicates the peak flux density histogram after correcting for the visibility function.
A preliminary version of these plots was shown in \citet{2015fers.confE..23G}.}
\label{visib}
\end{figure}

To construct the mosaic, the images of the seven pointings were transferred
back into {\sc aips}. We re-gridded the images using the task {\sc hgeom} to a
common centre, and produced a noise image for each pointing using the task {\sc rmsd}. 
The re-gridded images were combined, using the noise images as weights, with the task {\sc wtsum}. 
Finally, a noise image of the mosaic was generated.
The r.m.s. noise is $\simeq 1$ $\mu$Jy in the inner regions, and increases with
distance from the centre of the mosaic.
The visibility function (Fig.\,\ref{visib}), calculated over the region used to extract the source catalogue 
(see \S\,\ref{catalogue} ), shows that about 50 percent of the mapped area is characterized by an r.m.s. noise 
lower than 3\,$\mu$Jy, and remains $\leq10\mu$Jy across the whole field: this makes our survey the most sensitive 
yet at 5.5\,GHz.

\section{Source catalogue}
\label{catalogue}

\begin{center}
\begin{table*}
\caption{The catalogue of sources detected above $5\sigma$ at 5.5 GHz. We include here only the first ten sources in R.A. order, 
the full version of the catalogue is available as online-only material.
Column 1 gives the name of the source, columns 2 to 5 list the right ascension, declination and the respective errors, 
column 6 is the signal-to-noise ratio, columns 7 \& 8 are the peak brightness and the total flux density with the respective 
errors.  
Finally, column 9 lists the deconvolved FWHM sizes (major axis and minor axis) in arcsec, and the position angle in degrees
(a value of -99 in this column indicates that the source is made of multiple components).
For the radio sources classified as unresolved, the total flux is set equal to the peak brightness and the
deconvolved sizes are set to zero.}
\label{tab_sample}
\begin{tabular}{lccccccccc}
\hline
Source Name     &    R.A.       &     Dec      &$\sigma_\alpha$& $\sigma_\delta$ & SNR &   S$_p$       &  S$_T$           & Size \& PA \\
                &   (J2000)     &    (J2000)   &   (arcsec)   &  (arcsec)           &     & ($\mu$Jy beam$^{-1}$)  & ($\mu$Jy)        & (arcsec$\times$arcsec, deg)      \\
\hline
J123557+621536  &12 35 57.94    &  62 15 36.83 & 0.19         &0.18            &   7.1 &  $52.7\pm 7.5$  &  $82.4\pm 11.7$  & $0.74\times 0.25$ 118 \\
J123601+621126  &12 36 01.80    &  62 11 26.41 & 0.20         &0.20            &   5.6 &  $28.7\pm 5.2$  &  $51.3\pm  9.3$  & $0.83\times 0.37$ ~47 \\
J123603+621110  &12 36 03.25    &  62 11 10.97 & 0.19         &0.19            &   7.0 &  $30.4\pm 4.3$  &  $54.9\pm  7.8$  & $0.91\times 0.25$ ~55 \\
J123606+620951  &12 36 06.61    &  62 09 51.13 & 0.18         &0.18            &  11.0 &  $54.4\pm 5.0$  &  $54.4\pm  4.9$  & $0.00\times 0.00$ ~~0 \\
J123606+621021  &12 36 06.83    &  62 10 21.44 & 0.20         &0.20            &   5.6 &  $25.7\pm 4.6$  &  $43.0\pm  7.7$  & $0.91\times 0.00$ ~53 \\
J123608+621035  &12 36 08.12    &  62 10 35.89 & 0.17         &0.17            &  32.9 &  $129.9\pm 4.2$ & $129.9\pm  4.2$  & $0.00\times 0.00$ ~~0 \\
J123609+621422  &12 36 09.71    &  62 14 22.16 & 0.20         &0.20            &   5.1 &  $16.7\pm 3.3$  &  $16.7\pm  4.5$  & $0.00\times 0.00$ ~~0 \\
J123617+621011  &12 36 17.08    &  62 10 11.32 & 0.18         &0.18            &  12.9 &  $40.3\pm 3.2$  &  $40.3\pm  3.4$  & $0.00\times 0.00$ ~~0 \\
J123617+621540  &12 36 17.55    &  62 15 40.76 & 0.17         &0.17            &  39.7 &  $122.6\pm 3.3$ & $122.6\pm  3.2$  & $0.00\times 0.00$ ~~0 \\
J123618+621550  &12 36 18.33    &  62 15 50.58 & 0.18         &0.18            &  14.6 &  $45.1\pm 3.1$  &  $61.4\pm  4.2$  & $0.46\times 0.42$ ~32 \\
\hline
\hline
\end{tabular}
\end{table*}
\end{center}

To identify a sample of sources above a given local signal-to-noise ratio (SNR)
threshold in the 5.5\,GHz mosaic, we followed the same approach already successfully tested by other 
radio surveys \citep[e.g.][]{2003A&A...403..857B}.
We employed the {\sc aips} task {\sc sad} on both the mosaic image and the noise image to obtain a catalogue of 
candidate sources above the threshold of $4.5\sigma$. We limited the source extraction and the following analysis 
to a circular region of radius 7 arcmin around the mosaic centre. For each selected source, {\sc sad} estimates 
the peak and total fluxes, and the position and size using a Gaussian fit. Since the Gaussian fit may provide 
unreliable and biased results for low SNR sources, a better estimate of the peak brightness and position was obtained 
with a simple cubic interpolation around the fitted position using {\sc maxfit} in {\sc aips}. Only the sources for which 
the ratio of peak brightness and the local noise was $\ge 5$ (i.e. those with SNR$\ge$5) were included in the final 
catalogue.
We found a total of 100 components that were visually checked to identify possible multiple 
components of a single radio source. Eight components were converted into 2 single radio sources.
In these cases, we ``collapsed'' all the multiple components in the catalogue to a single source entry at the position 
of the brightest component, and we derived the total flux density integrating the brightness distribution 
over the area occupied by the source.
For all the remaining sources
we assessed the reliability of the total flux densities derived by {\sc sad} using simulated sources,  
added to the uv-dataset of the central pointing. The dataset, including the mock sources, was imaged and primary 
beam corrected as the original dataset. Forty mock sources, with total flux density in the range 20-200\,$\mu$Jy were 
inserted for each run of simulation in a region within a radius of 90\,arcsec from the pointing centre. 
Half of the inserted sources were point-like, while the remaining half had intrinsic sizes between 0.2 and 0.8\,arcsec. 
This procedure was repeated five times yielding to a sample of 200 mock sources. 
The total flux density recovered by {\sc sad} (1-component Gaussian fit) for each mock source was compared to the 
intrinsic one injected in the dataset. We found that the total flux densities derived with this procedure are 
systematically higher (on average by 15-20 percent) than the true values: the median, mean and standard deviation of the 
ratio between the measured and injected total flux using a simple 1-component Gaussian fit are 1.15, 1.20 and 0.25, respectively.
Therefore, we decided to manually fit each of our mock sources with a 2-component fit, including a Gaussian component 
and a zero-level baseline contribution.
The total flux densities obtained from these fits are in much better agreement with the true, injected values: 
the median, mean and standard deviation of the ratio between the measured and injected total flux densities 
using the 2-component (Gaussian + baseline) fit are 0.98, 1.02 and 0.15, respectively.
Summarising, for each single component source the peak brightness is measured with {\sc maxfit}, and the total 
flux density and sizes with the 2-component Gaussian fit.

\begin{figure}
\centering
\includegraphics[width=8cm]{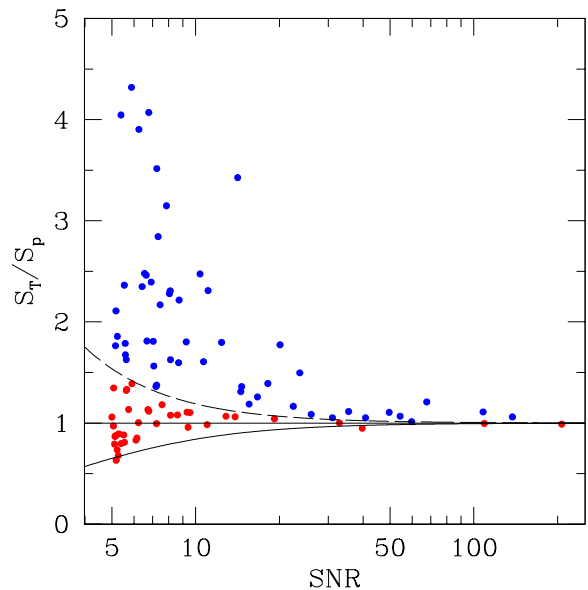}
\caption[]{Ratio between S$_{\rm t} $ and S$_{\rm p}$ as a function of the local SNR. 
Sources below the dashed line (red points) are considered unresolved,
while those above (blue points) are considered resolved (see text for details).
The horizontal solid line is drawn at $S_{\rm t} / S_{\rm p} =1$.
The lower solid-line envelope contains 90 percent of the sources with $S_{\rm t} / S_{\rm p} <1$
}
\label{flux_dist}
\end{figure}
The final catalogue contains 94 sources with SNR $\ge 5$. Table\,\ref{tab_sample} lists the source name, position in R.A. 
and Dec. with errors, signal-to-noise ratio, peak flux and total flux density with errors, and deconvolved sizes.
Sources that were classified as unresolved using the distribution of the total to peak flux ratio (see \S\,\ref{resolved}) 
have their total flux set equal to the peak flux and deconvolved sizes set to zero. The full version of 
Table\,\ref{tab_sample} is available as online-only material. Since the formal relative errors determined by 
Gaussian fits are generally smaller than the true uncertainties of the source parameters, we used the 
\citet{1997PASP..109..166C} error propagation equations to estimate the true errors on fluxes and positions 
\citep[e.g.][]{2000A&AS..146...41P, 2003A&A...403..857B}.
The contour plots of the 94 radio sources are shown in the Appendix (online material only).

The peak brightness ($S_p$) and total flux density ($S_t$) distributions for the 94 sources in our sample are shown in 
Fig.\,\ref{visib}, together with the expected peak flux density distribution corrected for local r.m.s. variations 
using the visibility function.

To test the reliability of lower SNR sources,
we quantified the number of possible spurious detections
due to random noise fluctuations and
associated to negative brightness peaks in the following way.
By assuming that negative and positive noise spikes
have a similar distribution in the 5.5\,GHz mosaic image,
we ran the task {\sc sad} on the negative mosaic map 
(i.e. the map multiplied by -1), with the same input parameters
used to extract the source catalogue.
We found 7 components with $\rm{SNR}\ge 5$
within the extraction area of the catalogue (7\,arcmin radius).
All these components are in the range $5\le \rm{SNR} < 5.5$.
In the radio catalogue there are 19 sources with $\rm{SNR}< 5.5$,
as shown by our analysis these lower SNR sources  
may be significantly contaminated by
false detection (7/19, $\sim$40 percent),
Since the following analysis is based on sources
with NIR identifications, we are confident that the fraction of spurious
radio sources is negligible, even at the lowest SNR values.

\subsection{Unresolved and extended sources}
\label{resolved}
Our source classification as unresolved or extended is based on the ratio between the total and
peak flux densities of the sources \citep[e.g.][]{2000A&AS..146...41P}:

\begin{equation}\label{eq-dec1}
S_{\rm t}/S_{\rm p}=(\theta_{\rm min} \; \theta_{\rm maj})
/(b_{\rm min} \; b_{\rm maj})
\end{equation}

where $\theta_{\rm min}$ and $\theta_{\rm maj}$ are the fitted source FWHM axes and
$b_{\rm min}$ and $b_{\rm maj}$ are the synthesized beam FWHM axes.
The flux density ratio distribution for all the 94 sources is shown
in Fig.~\ref{flux_dist}. As expected, at low SNRs we have sources with 
S$_{\rm t}$/S$_{\rm p}< 1$: this is due to the r.m.s. statistical errors which affect
our source size and, in turn, the flux density estimates.

To distinguish between unresolved and extended sources,
we derived the lower envelope of the flux ratio
distribution in Fig.\,\ref{flux_dist}
by fitting a curve above which there are at least 
90 percent (to discard possible outliers) of the sources with S$_{\rm T}$/S$_{\rm P}< 1$:, and then 
mirrored  it above the S$_{\rm T}$/ S$_{\rm P}$=1 value. 
The sources located above the upper envelope are considered extended,
while those below it are considered unresolved.
We stress that the total flux density of the sources is calculated
through a 2 component-fit (source $+$ background) which has proved to
be more reliable in measuring the total flux.

The upper curve can be described by the equation:
\begin{equation}\label{eq-dec2}
S_{\rm t} /S_{\rm p} = 1 +
\left[ \frac{ a }{ (S_{\rm p}/\sigma)^{\beta}}\right]
\end{equation}
where $a=6$ and $\beta=1.5$.
With this criterion, 56 (38) sources are considered resolved (unresolved).
We considered as reliable only the deconvolved angular sizes of the resolved sources,
while those of unresolved sources are set to zero in the catalogue
(see Table\,\ref{tab_sample}).

\subsection{Radio sources detected at 5.5\,GHz without a 1.4\,GHz counterpart}

The deepest observations at 1.4\,GHz of the GOODS-N field, published by \citet{2010ApJS..188..178M},
have a circular beam size of $\sim 1.7$\,arcsec and r.m.s noise level of $\sim 3.9$ $\mu$Jy\,beam$^{-1}$.
The 1.4\,GHz catalogue lists more than 1200 radio sources above a $5\sigma$ threshold of $\sim 20$
 $\mu$Jy\,beam$^{-1}$ at the field centre, within a region of $40\times 40$ arcmin$^2$. 
A significant fraction of the sources detected at 5.5\,GHz (17/94, 18 percent) have no
counterpart in the 1.4\,GHz catalogue. 
About half (9/17) of the sources without a 1.4\,GHz counterpart have SNR$>5.5$ or have a  NIR
identification (see \S\,\ref{id}) and these are the most reliable sources. 
These sources have upper limits in the
spectral index ranging from $\alpha < 0.66$ to $\alpha < -0.5$ (adopting the spectral index definition
$S\propto \nu^{-\alpha}$),  and will be discussed in more details in Paper II (Guidetti et al. in preparation). 
The number of the remaining sources (8/17) is consistent with the number of the expected spurious sources 
derived in \S\,\ref{catalogue}.

\section{Polarisation properties}

\begin{figure*}
\centering
\includegraphics[width=14cm]{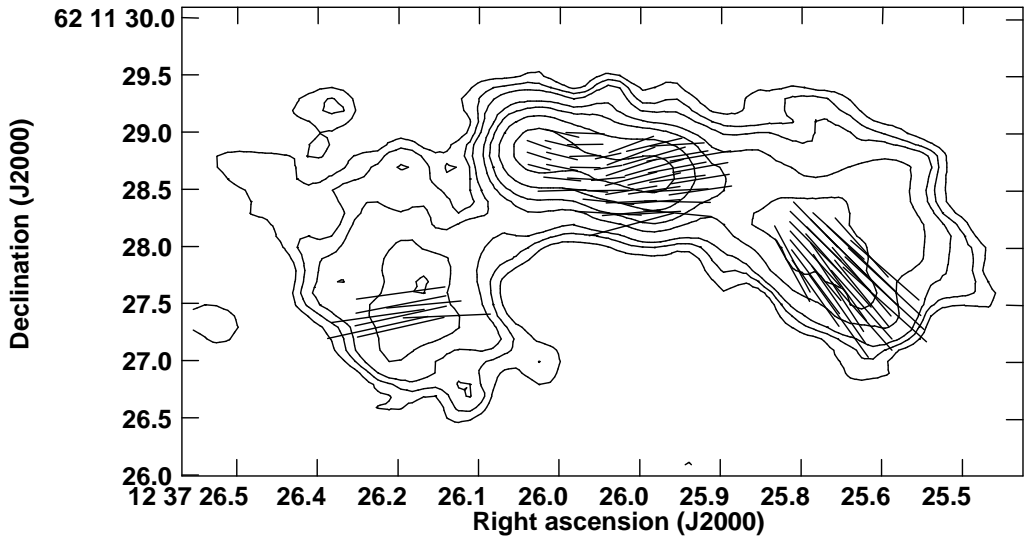}
\caption[]{Polarisation vectors with directions along the apparent electric field and lengths proportional
to the degree of polarisation at 5.5\,GHz, superimposed on the radio contours of total intensity
across J123726+621128 at the same frequency. The angular resolution is $\simeq 0.5$\,arcsec FWHM.
}
\label{fig-pol}
\end{figure*}

The mosaics of the Stokes parameters Q and U were imaged and assembled using the same
method applied to derive the total intensity mosaic. Noise images were also
derived.
The Stokes Q and U mosaics were combined to derive the polarised intensity
mosaic using the task {\sc comb} in {\sc aips}. The noise images were used to
clip signals below a threshold of $3\sigma$ in the polarised intensity image.
We then searched for polarised emission at the
positions of the sources in our sample.
We detected polarised emission in only eight sources. These are
listed in Table\,\ref{tab_pol}.
For each source we give the peak in polarised emission and its SNR, the total
polarised flux and the average fractional polarisation (calculated as the
total polarised flux divided by the total flux density of the source from Table\,\ref{tab_sample}).
Only two sources in Table\,\ref{tab_pol} show extended polarised emission.
In particular J123726+621128 has a wide-angle tail (WAT) morphology with polarisation detected in the
twin jets and in both
lobes as shown in Fig.\,\ref{fig-pol}.
The second extended, polarised source, J123644+621133, represents
the other galaxy in this field showing the classical FRI structure
(core+jets).

We estimated the bandwidth effects on the polarised emission of the sources,
assuming a rotation measure (RM) in the range 10-100\,radm$^{-2}$.
These are plausible values for the integrated RM of GOODS-N sources:
typical intrinsic RM values for extragalactic radio sources are in the range from a few radm$^{-2}$
in the poorest environments, through intermediate values of 30-100\,radm$^{-2}$, up to thousands
of radm$^{-2}$ in the centres of cool core clusters \citep{2011MNRAS.413.2525G, 2012MNRAS.423.1335G}.
The high Galactic latitude of the GOODS-N field ensures
a small contribution from the Galactic foreground.
For the worst case (i.e.
at the lowest frequency of our observations, 4.5\,GHz), the average rotation
across the bandwidth is $\sim$10\,degrees,
resulting in a depolarisation of 0.017 which is negligible compared to
the errors due to noise.

\begin{table}

\caption{Polarised sources. Col.\,1: Source name.
Col.\,2: Peak of polarised emission. Col.\,3: SNR of the polarised emission. Col\,4:
Total polarised emission. Col.\,5 Average fractional polarisation. 
}
\begin{center}
\begin{tabular}{lcccccc}
\hline
Source Name     & P       & SNR$_P$ & P$_{\rm tot}$ &  Pol. \\
                &($\mu$Jy)&         &($\mu$Jy)    &  (\%) \\
\hline
 J123623+621642 &  10.4   &  3.3    &  10.4       &  4.9  \\
 J123642+621545 &   6.0   &  3.3    &   6.0       & 10.8  \\
 J123644+621133 &  20.8   & 12.6    &  39.9       &  4.8  \\
 J123646+621629 &  10.9   &  5.0    &  10.9       &  7.5  \\
 J123700+620909 &   7.3   &  3.1    &   7.3       &  7.2  \\
 J123714+620823 &  19.5   &  4.8    &  19.5       &  1.0  \\
 J123721+621129 &  10.1   &  4.2    &  10.1       &  2.6  \\
 J123726+621128 &  28.4   &  9.4    &  86.8       &  8.2  \\
\hline
\end{tabular}
\end{center}
\label{tab_pol}
\end{table}

\citet{2014ApJ...785...45R} published the results of a polarisation
survey of the GOODS-N field at 1.5\,GHz. The observations had
a detection threshold of 14.5 $\mu$Jy in polarised emission and cover a
much larger area than that covered by our 5.5\,GHz data.
They detected 14 radio sources; two are in
the field of view covered by our observations: J123644.3+621132 and
J123725.9+621128 (wrongly named as J1237744.1+621128 in Table 1 in
 \citet{2014ApJ...785...45R}) and both show polarised emission at 5.5\,GHz.
These are the two extended sources discussed above, and the sources with the strongest polarised emission at 5.5\,GHz.
Taking into account that the synthesized beam of the 1.5\,GHz observations is
$1.6\times 1.6$ arcsec$^2$ (compared to $0.56\times 0.47$ arcsec$^2$ of our 5.5\,GHz observations),
the amount of polarised flux detected at the two frequencies is consistent.
Of the remaining sources we detect at 5.5\,GHz, only  J123714+620823
could have enough polarised flux to be detected at 1.5\,GHz.

\section{Near-Infrared identifications}
\label{id}

We searched for counterparts of the 5.5\,GHz sources in the ultra-deep $K_s$-band catalogue
of \citet{2010ApJS..187..251W}. The $K_s$-band imaging was performed with the Wide-field InfraRed Camera
(WIRCam) on the 3.6 m Canada-France-Hawaii Telescope (CFHT) and covers an area of 0.25 deg$^2$
with a $5\sigma$ point-source depth of $K_{s,AB}=24.45$ mag.
The field imaged at 5.5\,GHz is entirely covered by the WIRCam observations.

\begin{figure}
\centering
\label{id_dist}
\includegraphics[width=8cm]{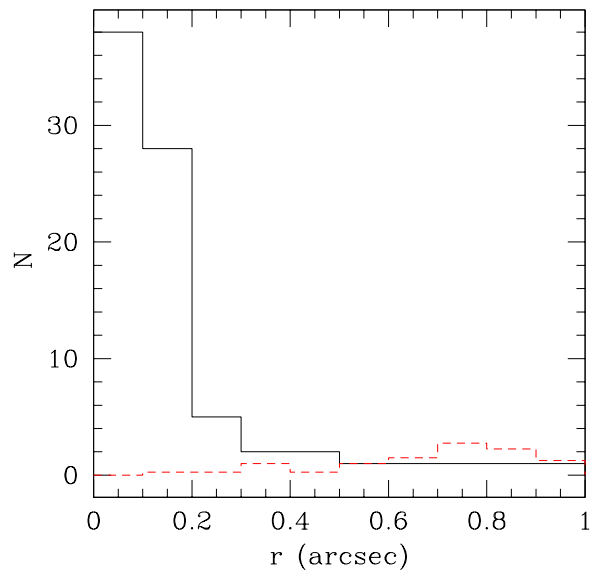}
\caption[]{Separation of the nearest K$_s$-band counterparts from the radio sources (solid black line),
and by shifting the radio sources by 1\,arcmin in four directions and
repeating the matching (dashed red line).}
\end{figure}

We initially searched for the nearest $K_s$-band counterpart within one arcsecond of each radio source: 
80 associations were found.
The distribution of the positional offsets between the radio sources
and the NIR nearest counterparts is shown in Figure\,\ref{id_dist}.
The majority of the counterparts are found within 0.5\,arcsec of the radio source positions.
The median shift in R.A. and Dec, considering all the counterparts within 0.5\,arcsec, is
$\Delta_{\rm R.A.}=-0.04\pm 0.10$\,arcsec and $\Delta_{\rm Dec}=+0.04\pm 0.05$\,arcsec, respectively.
These shifts are less than 1/10 of the radio synthesized beam and smaller than the errors 
(as estimated from the median absolute deviations); indicating that there is no evidence for any 
measurable systematic offset between the radio and NIR positions.

We checked for possible random coincidences by shifting the positions of the radio sources in the catalogue 
and searching for the nearest K$_s$-band counterparts,
obtaining the distribution of the separation for random coincidences (dashed line in Fig.\,\ref{id_dist}).
Based on the sky density of the NIR objects the probability of having a false association 
with a separation $\ge 0.5$ arcsec is $\sim 50$ percent.
Assuming $r=0.5$\,arcsec as the cut-off separation for a real identification, we
have 75 single identifications and expect two random coincidences.
A total of 19 sources have no NIR identification with $r\le0.5$\,arcsec.
We examined each of these sources and found that three are extended sources where the
radio peak is offset with respect to the possible optical counterpart, but the radio morphology
suggests that the optical association is correct,
resulting in 78/94
radio sources with a NIR counterpart in \citet{2010ApJS..187..251W}.

For the remaining unidentified objects we searched the catalogues of the Subaru
Multi Objects InfraRed Camera and Spectrograph (MOIRCS) ultra-deep survey
\citep{2011PASJ...63S.379K} and of the Cosmic Assembly Near-infrared Deep
Extragalactic Legacy Survey \citep[CANDELS][]{2014ApJS..214...24S} 
with 5 $\sigma$ depth of $K_{s,Vega}=24.1$
and $K_{s,AB}=24.7$, respectively. 
We found 5 new NIR counterparts
within 0.5\,arcsec. The level of  random
coincidences with these further NIR catalogues is still $\lsim 3$ percent.

In summary, we have a secure identifications for 83 radio sources
(88 percent of the whole radio catalogue). In the following discussion we will refer to these
83 sources as the NIR-identified sample.
If we restrict the radio sample to sources with SNR $\ge 5.5$,
then we have 79 radio sources, 76 of which have a secure NIR counterpart (96 percent of the radio sample).
The high fraction ($\sim 90$ percent) of reliable identifications is a natural consequence
of the depth of the NIR catalogues used to cross identify the radio sources, and is consistent
with that found in other studies \citep[e.g.][]{2006MNRAS.372..741S, 2017arXiv170309719S}.

\begin{figure}
\centering
\includegraphics[width=8cm]{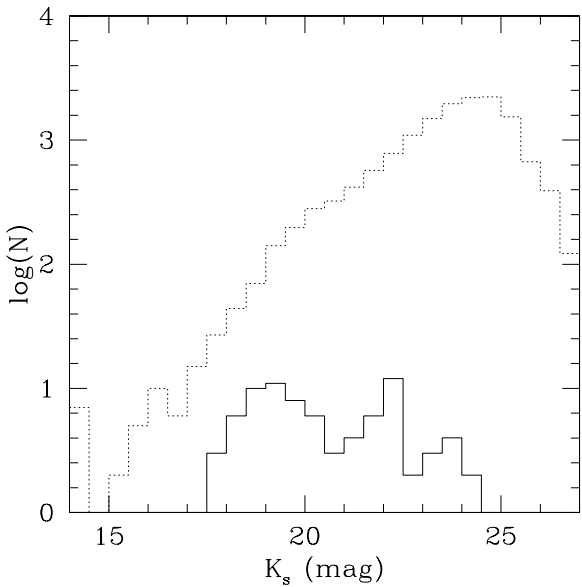}
\caption[]{$K_s$ magnitude distribution for the radio source counterparts (solid line) and
for the NIR-selected sample \citep{2010ApJS..187..251W} limited to sources within the same area covered by the
radio mosaic (dashed line).
}
\label{ks}
\end{figure}

The distribution of the $K_s$ magnitude for the NIR-identified sample is shown in
Fig.\,\ref{ks},
together with that of the \citet{2010ApJS..187..251W} $K_s$ selected sample, restricted to sources
in the area of our radio observations (dotted histogram).
The $K_s$ magnitude histogram of radio sources
displays a much flatter distribution
than the overall NIR sample, this should be due to 
our radio/NIR selection function and demonstrating that we are probing
a different source population than purely NIR-selected samples.

\subsection{Sources without near-IR counterparts}
Eleven sources are still unidentified at a limiting $K_s$ magnitude of 24.5-25.0. All of them are low
SNR radio sources (SNR $< 5.8$) and eight of them have SNR $< 5.5$. Moreover, none of the 11 unidentified sources
has a counterpart at 1.4 GHz in \citet{2010ApJS..188..178M}.
We searched for possible Infrared Array Camera (IRAC) and Multi-band Imaging Photometer for Spitzer (MIPS)
counterparts of the NIR unidentified radio sources in the
S-CANDELS catalogue \citep{2015ApJS..218...33A} and GOODS-N Legacy Survey 24\,$\mu$m catalogue \citep{2011A&A...528A..35M}.
We used a matching radius of 2 and 4\,arcsec for searching, respectively,  IRAC and 24\,$\mu$m MIPS counterparts.
One source is identified in both the IRAC and MIPS catalogues, for two radio sources an association is found in the
IRAC catalogue and, finally, for one source a counterpart is found in the  24 $\mu$m MIPS catalogue only.
In summary, we found four possible MIR counterparts among the eleven source without NIR
identifications.
One of the remaining sources
is actually a famous galaxy, being associated to
HDF\,850.1, the brightest submillimiter source in the field
\citep[][and references within]{2012Natur.486..233W}.

These radio sources, without a deep NIR identification but  with a MIR
or sub-mm counterpart, are potentially very interesting objects and we will
investigate their properties in a later publication, but since they do not
possess detections in other bands we cannot include them in the analysis described 
in \S\,\ref{agn}, which is limited to the NIR-identified sample.

\subsection{Spectroscopic \& photometric redshifts}

Many papers presents spectroscopic and/or photometric redshifts obtained in the
GOODS-N field \citep[e.g.][]{2001ApJ...551L...9C, 2004AJ....127.3121W, 2008ApJ...689..687B,
2011PASJ...63S.379K, 2014ApJS..214...24S, 2016ApJS..225...27M}.

In order to obtain a homogeneous and updated set of redshifts we adopt,
when available, the
redshifts from the 3D-HST Treasury survey catalogues \citep{2014ApJS..214...24S, 2016ApJS..225...27M}.
GOODS-N is one of the five CANDELS fields for which WFC3 G141 spectroscopic data are available.
Spectroscopic redshifts are either measured with space-based dispersion grisms \citep{2016ApJS..225...27M}
or obtained from ground-based  slit spectroscopy from the literature \citep{2014ApJS..214...24S}. Photometric
redshifts are determined with the EAZY code \citep{2008ApJ...686.1503B} and listed in \citet{2014ApJS..214...24S}.
The normalized median absolute deviation (MAD) scatter between the photometric and spectroscopic redshifts is
$\sigma_{\rm NMAD}= 1.48\times {\rm MAD} < 0.027\times (1+z)$ \citep{2014ApJS..214...24S}.
For five sources, not included in the 3D-HST photometric
catalogue of GOODS-N we searched the literature for an appropriate redshift.

Spectroscopic redshifts are available for 51 NIR counterparts and photometric redshifts  for 28 sources
yielding to $95$ percent (79/83) the fraction of NIR identified
sources with a redshift.
There are 4/83 NIR identified radio sources with no redshift measurement.
These are all located at the edges of the GOODS-N radio field in a region not covered by the IRAC
observations.

\begin{figure}
\centering
\includegraphics[width=8.7cm]{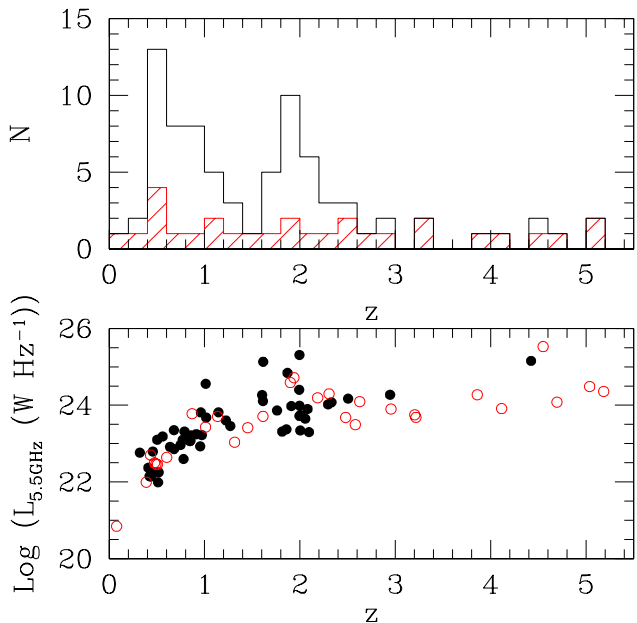}
\caption[]{Top: Redshift distribution (solid black line) for the 79 sources with known redshift 
(51 spectroscopic and 28 photometric), and for the 28 sources with only photometric redshift (hatched red line).
Bottom: Isotropic 5.5\,GHz luminosity as a function of redshift for the 79 sources 
with known redshift, spectroscopic (black filled points) or photometric (red empty points).
A fixed spectral index of $\alpha=0.7$ is used to convert flux densities to radio luminosities.} 
\label{z_dist}
\end{figure}

Hereafter we use spectroscopic redshifts where available, and photometric redshifts otherwise.
The redshift distribution is shown in Fig.\,\ref{z_dist}.
It appears to follow a bimodal distribution
which shows a correspondence to the (less prominent) peaks noted in the K$_s$
magnitude distribution.
There is a peak around $z\simeq 0.5$ with a tail extending to $z\simeq 1$ populated by the brighter sources
($17 < K_s < 21$) and a secondary peak around $z\simeq 2$.
One radio source is identified with a local galaxy at $z<0.1$.

The $z$-fitting procedure used in \citet{2016ApJS..225...27M}, based on combining grism and multi-band 
photometric datasets, provides photometric redshifts for some low redshift ($z\lsim 0.7$),
and often red and/or faint galaxies, although they might have a spectroscopic redshift in the literature.
For these objects, the photometric fit alone provides a more accurate redshift, and  the contribution
of the grism spectrum to the combined fit is negligible.
This explains the photometric redshifts assigned to some low redshift ($z\lsim 0.7$) sources in
Fig.\,\ref{z_dist}. It is worth noting that in all these cases
the photometric redshifts reported in \citet{2016ApJS..225...27M} are consistent with the spectroscopic 
values that are listed in other catalogues.

The median redshift for the 79 NIR identified radio sources with redshift information is $z=1.32$
($z=1.02$ for those with only spectroscopic redshifts), 8 radio sources  have a spectroscopic (1) or
photometric (7) redshift larger than 3.

The isotropic intrinsic radio luminosities at 5.5\,GHz (L$_{5.5{\rm GHz}}$)
for the sources with known redshifts are shown in Fig.\,\ref{z_dist}, where we assumed 
a radio spectral index $\alpha=0.7$  for the K-correction.
The median L$_{5.5{\rm GHz}}$ of the sources with redshift is
$4.4\times 10^{23}$ W\,Hz$^{-1}$.

\section{The AGN content of the 5.5\,GHz radio sample}
\label{agn}

In this section we use the exquisite multi-wavelength ancillary catalogues
of the GOODS-N field to identify systems  in the 5.5\,GHz radio sample that are likely hosting
an AGN.

Multi-wavelength observations have led to the identification 
of a two-fold mode of nuclear activity, RI-AGN and RE-AGN (see \S 1 ), 
which may reflect two different types of SMBH accretion and feedback.
RI-AGN and RE-AGN are also
named as ``radio-'' and ``quasar-'' mode AGN, respectively.

A full review of these two AGN populations and their properties
can be found in \citet{2014ARA&A..52..589H}.
In brief, RI-AGN stand out in the radio band,  without
accretion-related X-ray, optical, or MIR emission
\citep[e.g.][]{2007MNRAS.376.1849H}, and
showing only low excitation emission lines \citep{1979MNRAS.188..111H}.
Such AGN are associated with very low nuclear accretion rates (Eddington fraction $\ll$1 percent,
\citealt{1994ApJ...428L..13N, 1995ApJ...444..231N})
involving hot gas from the halo's atmosphere and
hosted in  massive galaxies within dense environments.
The feedback is based on the presence of
powerful radio jets that mechanically transfer the AGN energy into
the surrounding environment. It is widely accepted that recurrent
radio-mode AGN activity is a fundamental component of the lifecycle of massive galaxies,
responsible for maintaining these sources as ``red and dead'' once they have migrated
on the red sequence \citep[e.g.][]{2006MNRAS.365...11C, 2006MNRAS.368L..67B}

In contrast, RE-AGN emit powerfully over a
wide range of the electromagnetic spectrum
(MIR to X-rays).  They are typically faint radio sources, although
a small fraction emit large scale, relativistic radio
jets. They are also characterized by the presence of high-excitation emission lines.
Quasar-mode AGN are associated with
radiatively efficient ($>1$ percent of the Eddington rate) optically thick, geometrically thin
accretion disks \citep{1973A&A....24..337S}, accreting cold gas via secular processes,
hosted by galaxies found in less dense environments, and often showing
ongoing star-formation \citep[][and references therein]{2014ARA&A..52..589H}.
AGN feedback may occur through high-velocity winds, outflows generated close to the AGN,
radiation pressure on dusty gas, or thermal heating \citep[e.g.][]{2005ARA&A..43..769V, 2012ARA&A..50..455F}.

To assess whether a galaxy is hosting an active nucleus,
we applied a number of AGN selection criteria at IR, X-ray and radio wavelengths.
Throughout this analysis we will adopt the nomenclature RE- and RI-AGN to refer to
these two distinct AGN populations, even if we 
identify objects in these two classes  not 
directly deriving the accretion efficiency, but rather 
on the basis of of their radio/IR/X-ray properties (IR diagnostic plots, radio-excess or X-ray luminosity).
This means that objects classified as RE- or RI-AGN might not necessarily
have two different accretion modes, but simply reflect the reliability of the combined
radio/IR/X-ray diagnostics. 

The origin of the radio emission and its link with an AGN- or SF-dominated
host galaxy is deferred to Paper II, here we focus on deriving the fraction
of our 5.5\,GHz radio sample that is dominated by an AGN in at least one of the radio, infrared
or X-ray bands.
We first analyse the AGN content separately for each of the IR, X-ray, radio bands, then
the final classification scheme is obtained by combining all the criteria as follows.
Radio sources identified as AGN by the IR or X-ray diagnostics
are classified as RE-AGN.
Among the remaining sources, the RI-AGN are identified as those radio sources
having MIR colours typical of red and passive galaxies, or those showing a radio-excess.
All other sources, those not identified as AGN hosts by any of the aforementioned criteria,
are classified as SFGs.

\subsection{AGN infrared colour diagnostics}
\label{IR}

IR colour-colour criteria, based on
surveys conducted with {\it Spitzer} and {\it WISE},
are currently used to separate AGN from star-forming
or quiescent galaxies
\citep[e.g.][]{2004ApJS..154..166L, 2005ApJ...631..163S, 2007AJ....133..186L,
2012ApJ...748..142D}.
These diagnostics are based on the evidence
of a prominent dip in the SED of SFGs between the 1.6-$\mu$m stellar bump
and the emission from star-formation-heated dust at longer wavelengths. On the other hand,
luminous AGN should have a monotonically increasing power law SED across the
IRAC bands \citep[e.g.][]{1979ApJ...230...79N}, a consequence of X-ray-to-UV
AGN radiation reprocessed to IR wavelengths by a dusty torus surrounding the
central region \citep{1989ApJ...347...29S, 1992ApJ...399L..23P}.
This scenario applies to RE-AGN
and throughout this paper, when we use the term AGN without further specification, we refer
to the class of radiatively efficient AGN.

In the following, we make use of four IR colour-colour diagrams developed in recent years
to distinguish between AGN- and star-formation-dominated sources.
The four IR criteria are those presented by \citet{2012ApJ...748..142D}, \citet{2012ApJ...754..120M},
\citet{2012ApJ...759..139K, 2013ApJ...763..123K}.
All of these provide low contamination diagnostics ($\sim 10$ percent) in separating
SFGs and AGN up to high redshift ($z\sim 4$), by taking into account
the redshift evolution of IR colours.
This is essential for a sample of sources
spanning a wide range of redshifts in order to  properly classify
high redshift objects.
\citet{ 2012ApJ...748..142D} redefined, in a more restricted way,
the IRAC colours-based AGN selection criteria previously
developed by \citet{2004ApJS..154..166L, 2007AJ....133..186L}
obtaining a highly-reliable classification
for deep IRAC surveys (referred to as the ``Donley wedge'' hereafter).
\citet{2012ApJ...754..120M} developed their own IR
colour-colour  diagram, using $K_s$ and IRAC bands (the KI diagram) and,
finally, \citet{2012ApJ...759..139K, 2013ApJ...763..123K} presented two different combinations of colours that
combine MIR and far-IR (FIR) photometry ({\it Spitzer} IRAC/MIPS and {\it Herschel} PACS/SPIRE)
to classify high redshift ($z=0.5-4$) galaxies selected at 24\,$\mu$m with
{\it Spitzer} IRS spectroscopy.
These criteria will be briefly presented in the following.
For a full discussion we refer to the
original papers listed above.

\subsection{IR classification of the 5.5\,GHz radio sources}
In \S\,\ref{ir_class} we present the results obtained applying the
IR colour-colour diagnostics to the radio selected sample in GOODS-N, adopting
the following nomenclature:

\begin{itemize}
\item those radio sources classified as AGN by only one diagnostic diagram 
are dubbed as RE-AGN-candidates, while RE-AGN are those classified as
AGN by at least two of the four IR colour-colour plots. 
\item Radio sources with MIR colours consistent with those of red passive galaxies 
are defined as RI-AGN.
Indeed, in these sources the AGN can be detected only in the radio band with no evidence
of accretion-related emission or recent star-formation in the IR or X-ray bands.
\item Galaxies which do not fit in the AGN regions of \textit{any} of the used 
IR colour-colour plots are classified as SF/hybrid systems (SF/hyb).
\end{itemize}
It is important to clearly state that the term SF/hyb is chosen to underline that some of these sources 
may not be necessarily pure SFGs, 
as most of the IR criteria here used  
are conservative at expenses of completeness.
Of course, the SF/hyb radio sources are not AGN-dominated in the IR, they could include purely SFGs,  
hybrid sources where AGN and star-formation coexist, or IR weak AGN.
In \S\,\ref{xray_class}, \S\,\ref{rx_class} and \S\,\ref{vlbi} we will check the SF/hyb radio sources 
for AGN-related radio emission using 
other diagnostics (X-ray luminosity, radio-excess, and compactness).

\subsection{MIR to FIR photometry}
The GOODS-North field has a wealth of ancillary information at IR wavelengths. In particular we used the
IRAC photometry (3.6, 4.5, 5.8, 8.0\,$\mu$m) reported in
\citet{2010ApJS..187..251W}, and MIPS photometry at 24\,$\mu$m from
\citet{2011A&A...528A..35M, 2013A&A...553A.132M},  both measured by {\it Spitzer}.
{\it Herschel} imaging covers the entire GOODS-N field with the Photoconductor Array
Camera and Spectrometer \citep[PACS, 100 and 160\,$\mu$m;][]{2010A&A...518L...2P}
and Spectral and Photometric Imaging Receiver \citep[SPIRE, 250, 350 and 500\,$\mu$m;][]{2010A&A...518L...3G}
data, as part of the PACS Evolutionary Probe \citep[PEP][]{2011A&A...532A..90L}
and the GOODS-{\it Herschel} \citep[GOODS-H][]{2011A&A...533A.119E}.
The FIR photometry is taken from the PEP DR1 catalogue \citep{2013A&A...553A.132M} and
the GOODS-H catalogue \citep{2011A&A...533A.119E}.
IRAC photometry in all four bands is available for 77/83 radio sources.
Four of the six missing sources lie just outside of the IRAC area coverage, while
the remaining two are not detected in any of the IRAC bands.

The number of radio sources with a detection in all of the four mid/far-IR bands
(250, 24, 8.0, 3.6\,$\mu$m) is 47. To these we add 19 sources for which we
use a $3\sigma$ upper limit for the 250$\mu$m flux. On the other hand, we have 52 radio sources detected in all the four
mid/far-IR bands  (100, 24, 8.0, 3.6\,$\mu$m)
plus 14 for which we use a $3\sigma$ upper limit
for the 100$\mu$m flux. In summary, the
sample of radio sources to which we can apply the mid/far-infrared colour-colour AGN selection
criteria contains 66 objects.

\begin{figure*}
\centering
\includegraphics[width=8cm]{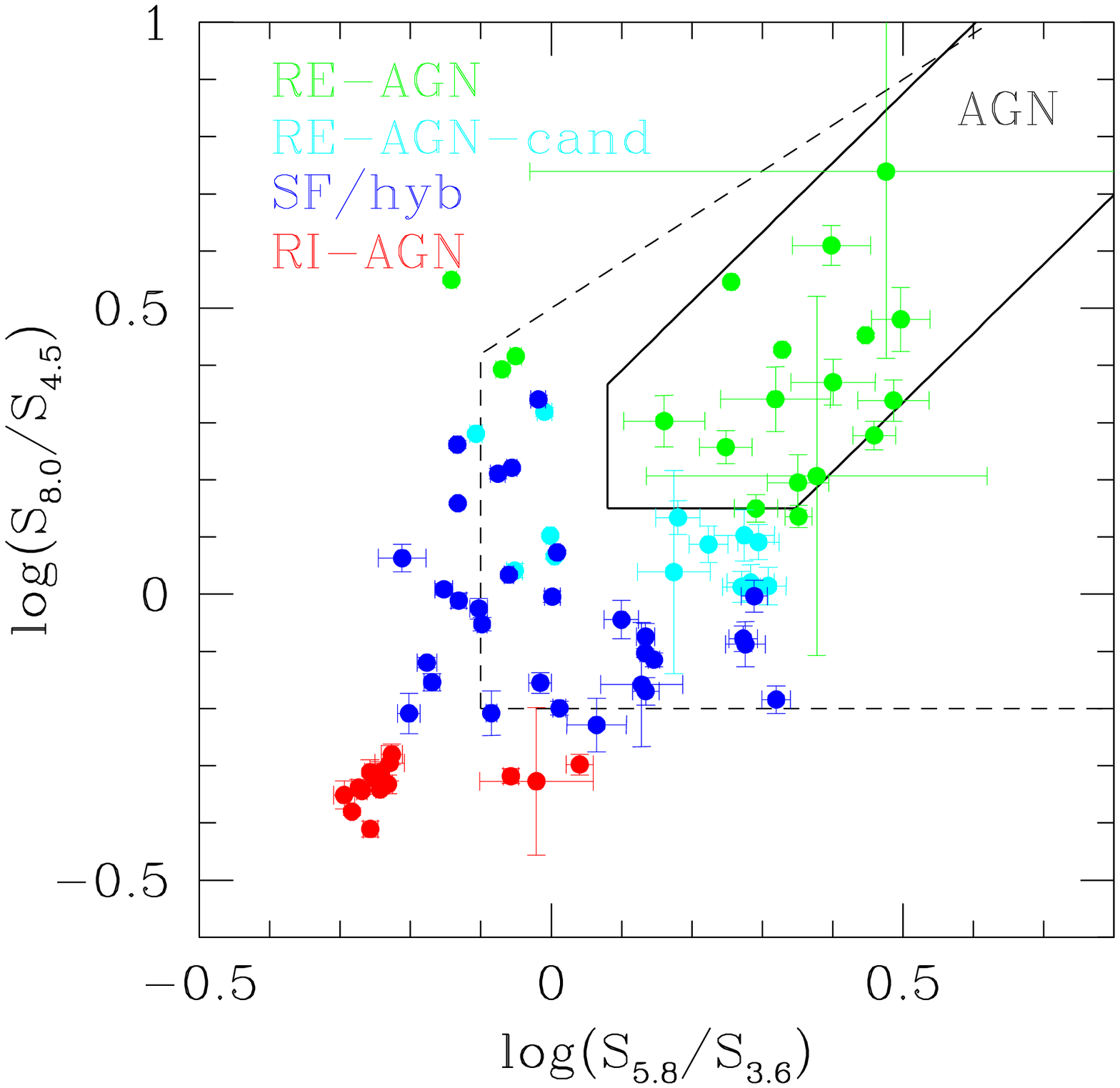}
\includegraphics[width=8cm]{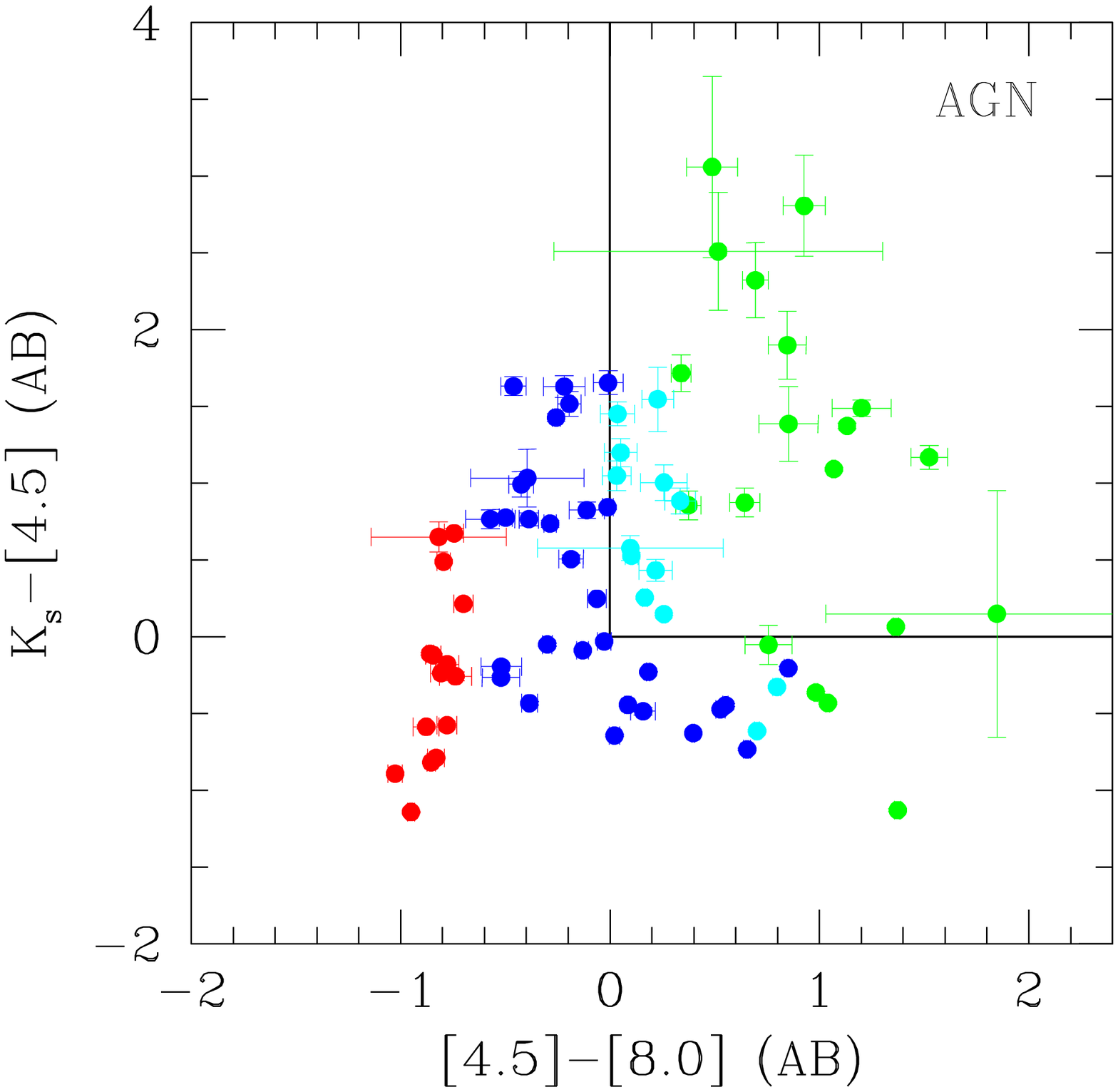}
\includegraphics[width=8.cm]{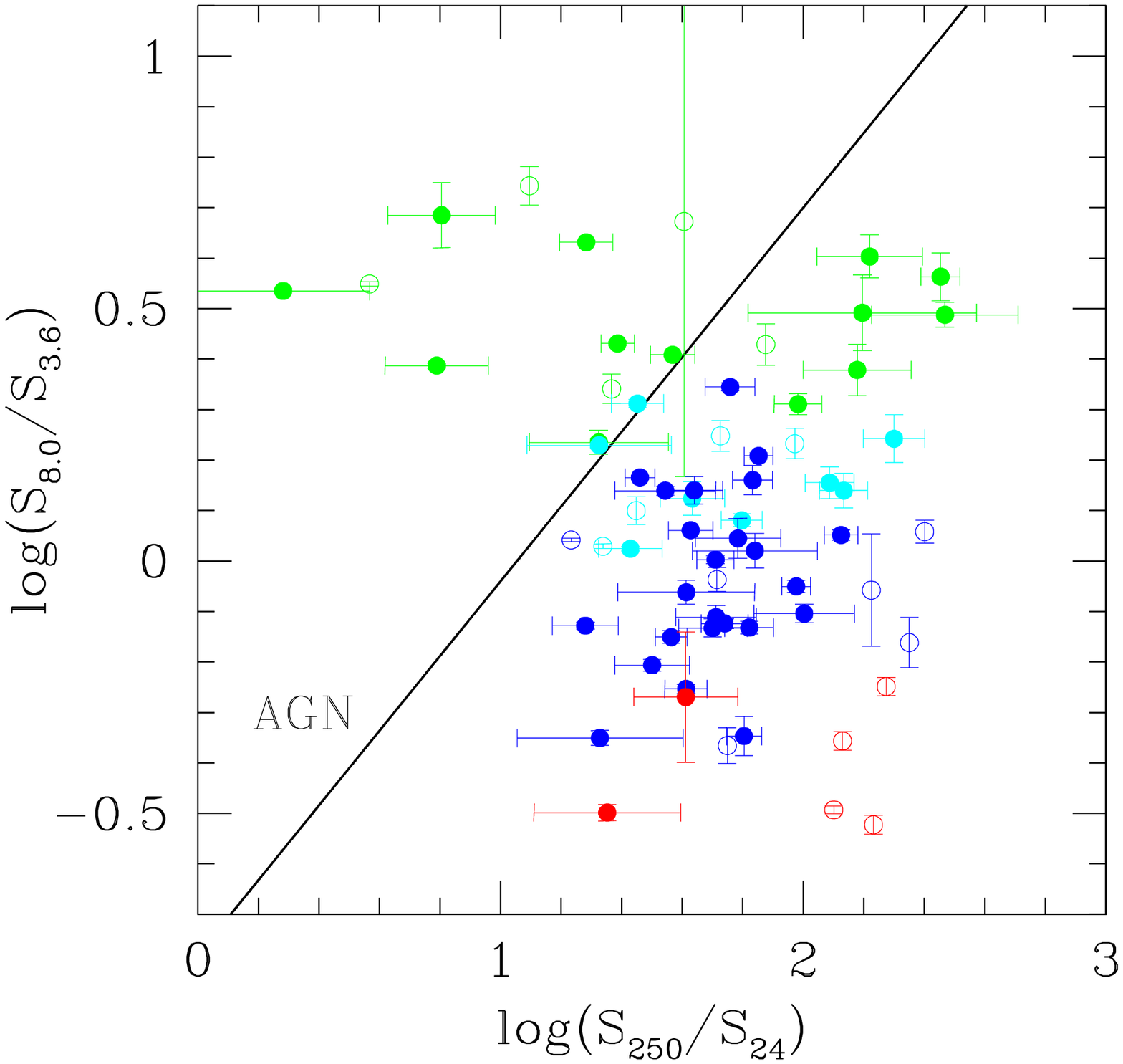}
\includegraphics[width=8.cm]{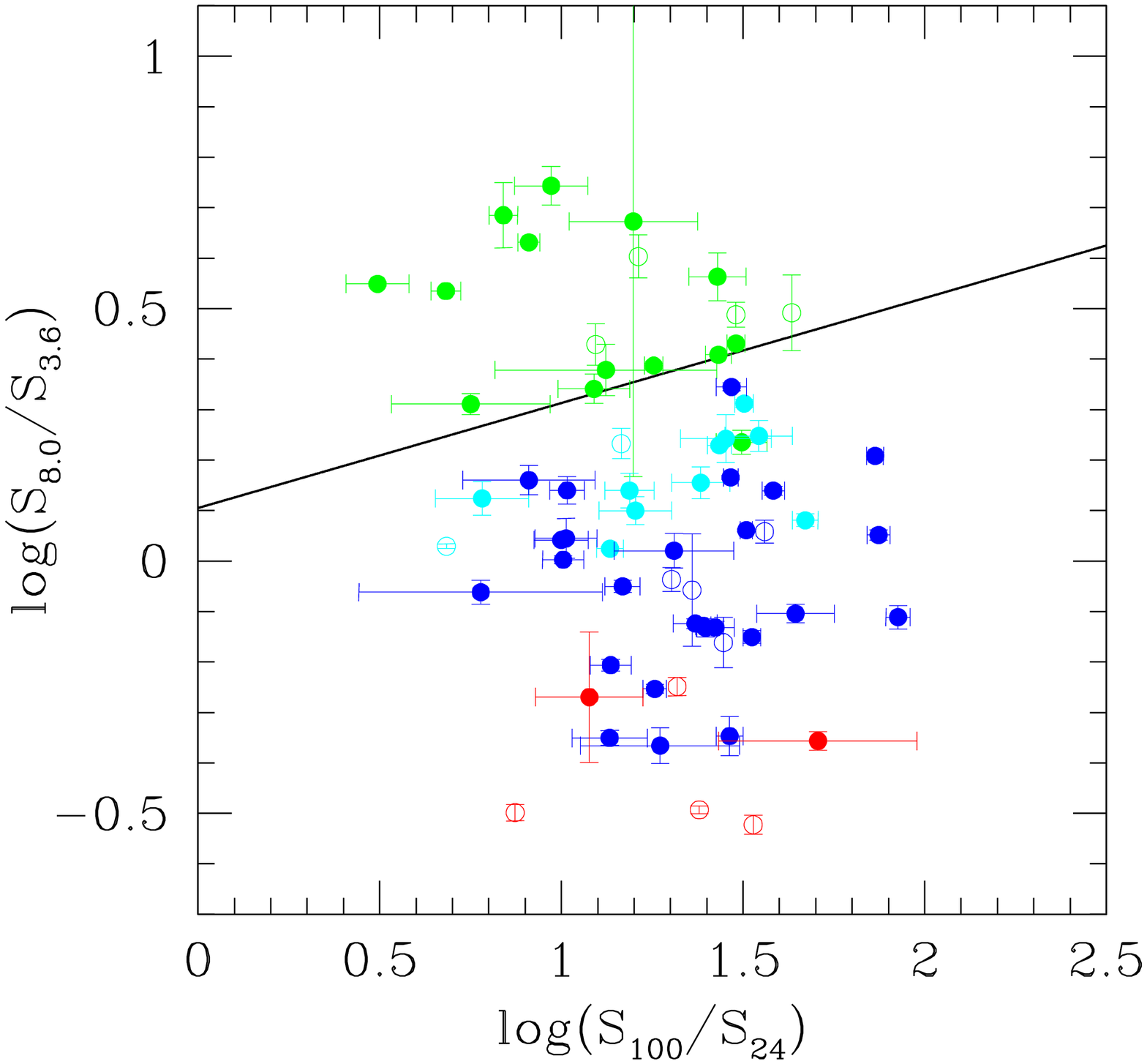}
\caption[]{The four IR colour-colour plots used to classify the 5.5\,GHz  radio sources in the GOODS-N
field. In each panel different colours are used to identify different classes of sources based on the classification
obtained by combining all the four IR plots: RE-AGN are shown in
green, RE-AGN-candidates in cyan, RI-AGN in red, and the remaining sources (SF/hyb
galaxies) in blue.
Top-Left: IRAC colour-colour plot, the dashed-line wedge shows the AGN selection region from \citet{2007AJ....133..186L},
 while the smaller area enclosed by a solid-line is the revised wedge from \citet{2012ApJ...748..142D}, 
 used in this paper.
Top-Right: KI colour-colour plot \citep{2012ApJ...754..120M}. The region populated by AGN is
delimited by the solid line
defined as $K_s-[4.5] > 0$ and $[4.5]-[8.0]>0$.
Magnitudes are in the AB photometric system.
Bottom-Left: IR colour-colour plot showing $\log(S_{8.0}/S_{3.6})$ versus $\log(S_{250}/S_{24})$.
The line separate RE-AGN (above the line) from star-forming or passive galaxies (below the line),
according to \citet{2012ApJ...759..139K, 2013ApJ...763..123K}
Bottom-Right: IR colour-colour plot showing $\log(S_{8.0}/S_{3.6})$ versus $\log(S_{100}/S_{24})$.
The line separates RE-AGN (above the line) from star-forming or passive galaxies (below the line),
according to \citet{2012ApJ...759..139K, 2013ApJ...763..123K}.
In the two lower plots, empty symbols refer to 100\,$\mu$m and 250\,$\mu$m upper limits.
}
 \label{midir1_plot}
\end{figure*}

\subsection{IR classification}
\label{ir_class}
Fig.\,\ref{midir1_plot} shows the four IR colour-colour plots used in this work.
The IRAC colour-colour diagnostic (top-left panel in Fig.\,\ref{midir1_plot}) can be applied to 77 of our 5.5\,GHz
selected sources, detected in all the IRAC bands.
The original AGN selection wedge is plotted
\citep{2007AJ....133..186L} as well as
the revised region assumed by \citet{2012ApJ...748..142D}.
Thirteen radio sources (about 17 percent) are inside the Donley wedge, and
all are power-law AGN: their flux densities are such that
$S_{3.6}<S_{4.5}<S_{5.8}<S_{8.0}$, within the photometric errors.
All of them are classified as RE-AGN since they are selected by at least two different IR criteria.
This confirms that AGN selected using the Donley wedge are highly reliable and not significantly affected
by contamination.
The drawback of this method is the lower level of completeness: only $68$ percent (13/19) of the RE-AGN
and $41$ percent (13/32) of the total number of RE-AGN and RE-AGN-candidates
in our 5.5\,GHz radio-selected sample are found inside the Donley wedge.

The RE-AGN (6 sources) and RE-AGN-candidates (13 sources) outside the Donley wedge are roughly evenly split
in two groups. Those with $z>$1.5  are found clustered in the region close to the wedge and bridging the
gap with the population of SF/hyb galaxies ($\log(S_{5.8}/S_{3.6})\simeq 0.25$ and
$\log(S_{8.0}/S_{4.5})\simeq 0.1$).
On the other hand, sources with $z<$1.5 span a wider range of $\log(S_{8.0}/S_{4.5})$ (indicating different 
levels of reddening) and are mixed to the SF/hyb galaxies.

A group of radio sources is closely clustered in the bottom-left of the diagram (top-left panel in Fig.\,\ref{midir1_plot}
).
These galaxies have MIR colours consistent with those expected for red and passive galaxies at $z\lsim 1$.
To allow for possible higher redshift ($z\simeq 2$) quiescent galaxies (see Fig.\,2 in \citet{2012ApJ...748..142D} for
the evolutionary tracks of passive galaxies), we consider all the 15 radio sources
with $\log(S_{5.8}/S_{3.6})<0.05$ and $\log(S_{8.0}/S_{4.5})< -0.25$
as radio-emitting RI-AGN hosted by red passive galaxies.

The KI criterion (top-right panel in Fig.\,\ref{midir1_plot})
identifies as AGN the sources with $K_s-[4.5] > 0$ and $[4.5]-[8.0]>0$, where AB magnitudes are used.
This method selects the largest number of AGN (15 RE-AGN and 11  RE-AGN-candidates) and
delivers the highest level of completeness both for the RE-AGN (88 percent, 15/19 sources) and the
overall RE-AGN and RE-AGN-candidates (81 percent, 26/32 sources).
However, given that 11 radio sources are identified as AGN-candidates only by this diagnostic, we must consider 
the possibility that some of these 11 sources could be misclassified SF/hyb galaxies.
We will return to this point later, we just note 
that the KI diagnostic is most effective in selecting AGN  at $z\simeq$ 2-3 \citep{2012ApJ...754..120M},
and  that 7/11 of the RE-AGN-candidates are indeed in this redshift range.

Before reviewing the results from the diagnostics using FIR photometry \citep{2012ApJ...759..139K, 2013ApJ...763..123K},
we note that this is possible for a smaller number
of sources (66 compared to 77 objects). Of the 11 sources without a FIR counterpart,
9 are associated with
RI-AGN hosted by red passive galaxies which are typically faint in the MIR, and
therefore mostly undetected at 24 $\mu$m.
The IRAC/{\it Herschel-250} ($\log(S_{8.0}/S_{3.6})$ vs. $\log(S_{250}/S_{24})$, lower-left panel in Fig.\,\ref{midir1_plot}),
selects only 58 percent (11/19) of the RE-AGN and 41 percent (13/32) of
RE-AGN plus RE-AGN-candidates.
We cannot exclude the possibility that a few sources classified as RE-AGN by at least two of the other criteria
could shift into the AGN region,
since they have only an upper limit for the {\it Herschel} flux.
The IRAC/{\it Herschel-100} ($\log(S_{8.0}/S_{3.6})$ vs. $\log(S_{100}/S_{24})$, lower-right panel in Fig.\,\ref{midir1_plot}), is
more effective in selecting RE-AGN (89 percent, 17/19) and  all the radio sources above the threshold
line are confirmed as RE-AGN by at least one other
criterion, implying a high reliability coupled with a high level of completeness.
The RE-AGN-candidates occupy an intermediate region between the RE-AGN and the population
of SF/hyb galaxies, while the few RI-AGN with a 24\,$\mu$m detection have
bluer $\log(S_{8.0}/S_{3.6})$ colours.

\begin{figure}
\centering
\includegraphics[width=8cm]{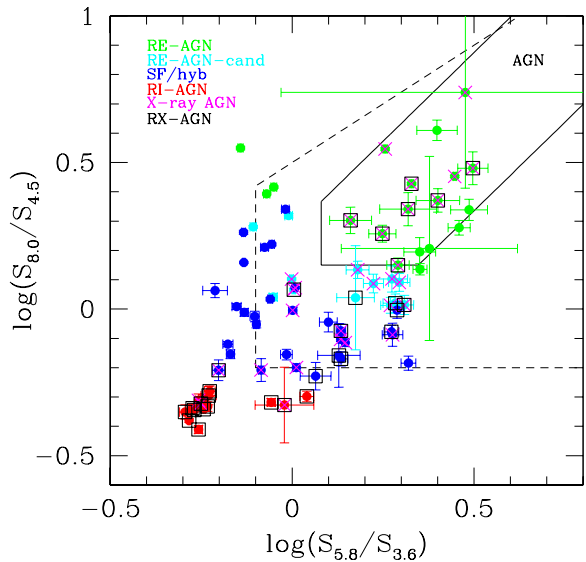}
\caption[]{The IRAC colour-colour diagram for the 77 5.5\,GHz selected radio sources detected at all IRAC bands, 
highlighting X-ray AGN (magenta crosses) and radio-excess sources (black squares).}
\label{donley2}
\end{figure}

We note that for the present sample of radio sources, the four IR diagnostics we have used
are practically equivalent to selecting sources in the IRAC colour-colour diagram with the cuts
$\log(S_{5.8}/S_{3.6}) > -0.1$ and $\log(S_{8.0}/S_{4.5}) >0$  with a $\simeq 10$ percent level of
incompleteness (1 RE-AGN and 2 RE-AGN-candidates are located outside this region)
and contamination (3 SF/hyb sources fall within this region).

In summary, using the four IR colour-colour diagnostics we find 19 RE-AGN,
13 RE-AGN-candidates,
and 15 RI-AGN associated with red passive galaxies.
Considering both RE-AGN and RE-AGN-candidates and the RI-AGN, about 61 percent (47/77) of the radio sources
with available infrared photometry are classified as AGN.
The remaining 30 sources, the SF/hyb systems, are not identified as AGN hosts by any
of the four IR criteria.
It is important to underline that we cannot exclude the presence of weak nuclear activity,
since the IR colour-colour plots
just tell us that the AGN, if present, does not dominate the IR emission (for the RE-AGN) or
has MIR colours not compatible with those of a red passive elliptical (for the RI-AGN).
So, these SF/hyb galaxies could be either pure SFGs, hybrid systems, or IR-weak AGN.

\subsection{X-ray AGN}
\label{xray_class}

We searched for X-ray bright AGN in our 5.5\,GHz catalogue, by using the 2\,Ms {\it Chandra} Deep Field-North
improved point-source catalogue \citep{2016ApJS..224...15X}, which covers the whole area of our radio observations.
The main catalogue lists 683 X-ray sources detected using {\tt WAVDETECT} with the following criteria: 1) a false positive
probability threshold of $10^{-5}$ in at least one of the three standard X-ray bands (full band, 0.2-7\,keV;
soft band, 0.2-2\,keV; hard band, 2-7\,keV); and 2) a binomial probability source-selection criterion of $P<0.004$
\citep{2016ApJS..224...15X}.
This new approach maximizes the number of reliable sources detected, yielding 196 main catalogue
new sources compared to \citet{2003AJ....126..539A}.
Using the same {\tt WAVDETECT} threshold but
 $0.004<P<0.1$, and limited to NIR-bright counterparts ($K_s< 22.9$\,mag), results in a supplementary catalogue of 72 additional X-ray sources.
\begin{figure}
\centering
\includegraphics[width=8.5cm]{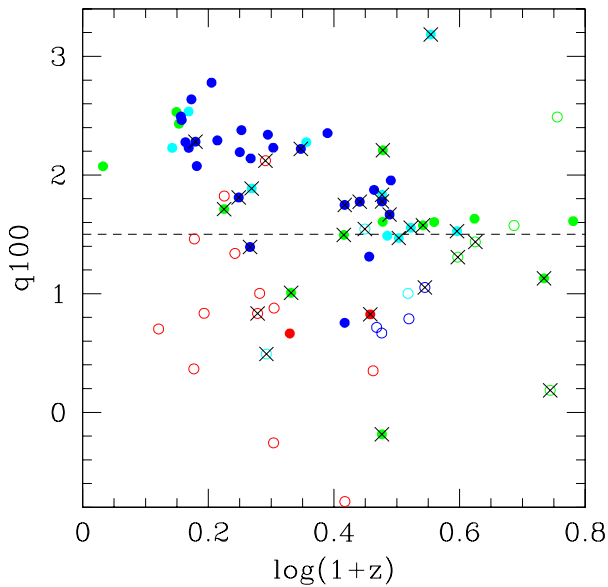}
\caption[]
{The $q_{100}=\log(S_{100\rm \mu m}/S_{\rm 1.4GHz})$ as a function of redshift for the 5.5\,GHz radio-selected sample
with full IRAC detection. Symbols are the same as in Fig.\ref{midir1_plot}, except that open symbols
are used for sources that have only an upper limit at 100\,$\mu$m.
}
\label{plotq100}
\end{figure}
The X-ray sources in the catalogue are already associated to a K$_s$-band counterpart using the catalogue of \citet{2010ApJS..187..251W}.
We find that 50 radio sources have an X-ray counterpart in the main catalogue, and two
radio sources have an association in the supplementary catalogue.
We checked for possible X-ray counterparts to the radio sources without a NIR identification, but
found no X-ray counterpart within 1\,arcsec.
The fraction of radio sources
with an X-ray counterpart is $\simeq 55$ percent (52/94), raising up to $\simeq 63$ percent (52/83) for NIR-identified 
radio sources. These fractions are higher than those found
in similar works on extragalactic radio sources \citep[e.g.][]{2013MNRAS.436.3759B}.
For the {\it Chandra} Deep Field-South, \citet{2013MNRAS.436.3759B}
exploited X-ray data sets twice as deep,  4\,Ms \citep{2011ApJS..195...10X} plus 250\,ksec observations 
\citep{2005ApJS..161...21L},  obtaining X-ray detections for
25 percent of their radio sources. Therefore, we reasonably conclude that the higher fraction of radio sources
with an X-ray association found in the 5.5\,GHz catalogue is not a consequence of the depth
of the X-ray data, but rather of the larger fraction of AGN sources in our catalogue. This is likely due
to the higher frequency (5.5 against 1.4\,GHz) and to the better angular
resolution (see \S\,\ref{disc}).

We classified our radio sources as X-ray AGN if
the observed hard-band 2-7 keV luminosity (not corrected for intrinsic absorption)
$L_{2-7 keV,obs} > 10^{42}$ erg\,s$^{-1}$, i.e. the typical X-ray luminosity threshold adopted to
separate AGN-related from star-formation-related X-ray emission.
If the hard-band X-ray luminosity is derived from an upper limit of the 2-7 keV flux,
we require a de-absorbed total-band luminosity $L_{0.2-7 keV,int}> 2\times 10^{42}$  erg\,s$^{-1}$.
On the basis of these two criteria, a total of 30 radio sources are classified
as X-ray AGN. These are marked with magenta crosses overlaid on the IR class symbols
in Fig.\ref{donley2}.
As already shown in many papers \citep[e.g][]{2008ApJ...680..130C}, in the IRAC colour-colour plot
the X-ray AGN tend to have colours consistent with a
power-law SED from the bluest to the reddest colours.
Only 3 bright X-ray sources are found in the region of the RI-AGN, which,
by definition, are X-ray weak.

We derived the median value of
the 2-7\,keV luminosity for the IR classified sources,
taking into account the upper limits by using the Kaplan-Meier
estimator and the code {\tt ASURV},
which implements the methods described by \citet{1985ApJ...293..192F} and \citet{1986ApJ...306..490I}
to properly handle censored data. Out of the 19 RE-AGN, 14 are
detected in the 2\,Ms catalogue with a median 2-7\,keV luminosity
(including 3 upper limits) of $2.4\times 10^{43}$ erg\,s$^{-1}$,
10 of these are X-ray AGN.
All the 5 RE-AGN undetected in the X-ray band are located close to the lower diagonal line
of the Donley wedge and have $z> 2.6$. Therefore, we can reasonably assume they are not detected in the 
2\,Ms X-ray image for sensitivity reasons, and otherwise they would have been classified as X-ray AGN due 
to their high redshift.
Out of the 13 RE-AGN candidates, 11 have a detection or upper limit in the hard-band, with 8 sources
that can be classified as X-ray AGN on the basis of the hard-band or total-band X-ray flux density.
The radio sources classified as RE-AGN candidates are typically
almost one order of magnitude less luminous than the RE-AGN
in the 2-7\,keV X-ray band: they have
a median X-ray luminosity (including 5 upper limits) of $3.1\times 10^{42}$ erg\,s$^{-1}$.
Out of the 15 RI-AGN associated with passive galaxies, 9 have a detection or upper limit in the hard-band,
and 3 are X-ray AGN.
The median  2-7\,keV X-ray luminosity of the 9 X-ray-detected RI-AGN (including 3 upper limits)
is $1.6\times 10^{41}$ erg\,s$^{-1}$.

In summary, 56 percent (18/32) of the radio selected RE-AGN (RE-AGN-candidates included)
are also X-ray AGN, compared to the 20 percent (3/15) of the RI-AGN.
It is interesting to note that we also find
X-ray bright sources in the remaining sources that were not classified
as AGN from any of the four IR colour-colour diagrams, the so-called SF/hyb sources.
Out of the 30 SF/hyb systems, 18 are detected in the X-ray, with 9 (30 percent) having
X-ray luminosities typical of AGN.
A fraction of the X-ray emission can be produced by star-formation processes
and, therefore, we might wrongly classify some of these sources as X-ray AGN.
To investigate this possibility, we applied two further tests to  
the 9 SF/hyb sources with X-ray luminosities typical of an AGN.
Firstly, we derived the X-ray to optical flux ratio, defined as
$\log(f_X/f_{\rm opt})=\log(f_{\rm 0.5-2keV}) + 0.4\times R + 5.71$, where R is the R-band magnitude.
Sources with $\log(f_X/f_{\rm opt})> -1$ are assumed to be powered by an AGN
\citep[e.g.][]{2002AJ....124.2351B, 2004AJ....128.2048B}. The X-ray to optical flux ratio is effective in separating
AGN and SFGs up to $z\sim2$ \citep{2014MNRAS.443.3728S}, and
the nine SF/hyb sources have redshifts in the range $z=0.5-2.5$, with only two objects with $z>2$ (2.08 and 2.5).
Then we compared the X-ray luminosities with those expected from
star-formation processes using the relation between X-ray luminosity and total IR luminosity
derived by \citet{2014MNRAS.443.3728S}.
The total IR luminosities of these nine sources were derived fitting the mid-to-far infrared
SED  using the IDL code developed by \citet{2012MNRAS.425.3094C}:
a simultaneous joint fit to a single dust temperature greybody in the FIR  
(for the cold dust component representing the reprocessed SF emission)
plus a MIR power law (for the hot dust component from AGN).
The total IR luminosity is derived from the greybody
component.

In just one case we find that the X-ray-to-optical ratio is consistent with that expected from a SFG
and the contribution of the star-formation to the X-ray luminosity is significant ($\sim 25$ percent).
Even correcting for this contribution, the hard-band X-ray luminosity of this source exceeds the threshold here used.
Therefore, we conclude that in all the SF/hyb X-ray powerful sources (with the possible exception of one source
that has mixed diagnostics) the X-ray emission is dominated
by an AGN component. All these nine sources, previously not classified as AGN on the basis
of the colour-colour diagnostics are X-ray bright AGN, confirming
that a significant fraction of radio-selected SF/hyb objects show nuclear activity
that is not detected by the IR colour-colour diagnostics.
We note that all the SF/hyb sources with $z>1.5$ are either undetected in the X-ray or, if detected, are classified
as AGN on the basis of their hard-band luminosity.
The median 2-7\,keV luminosity of the 18 SF/hyb detected in the full band X-ray image
(including 7 hard-band upper limits) is
$4.1\times 10^{41}$ erg\,s$^{-1}$.

\subsection{Radio-excess sources}
\label{rx_class}

The correlation between the total IR luminosity and the 1.4\,GHz radio luminosity
for galaxies with ongoing star-formation is one of the
tightest in astrophysics.
The relationship holds over a very wide range of redshifts and
luminosities, from normal, radio-quiet spirals to ultra-luminous IR galaxies (ULIRGs)
\citep{2001ApJ...554..803Y, 2002A&A...384L..19G, 2004ApJS..154..147A},
and is one of the most useful diagnostic tools in revealing excess
radio emission exceeding that
expected from pure star-formation processes.

Several studies have replaced the FIR flux with the monochromatic flux at 24\,$\mu$m
\citep[e.g.][]{2013MNRAS.436.3759B} or at 100\,$\mu$m \citep[e.g.][]{2013A&A...549A..59D}
as a proxy for the FIR emission.
In particular, \citet{2013A&A...549A..59D} showed that a simple cut at $q_{100}<1.5$
(where $q_{100}=\log(S_{100\mu m}/S_{1.4\rm GHz})$) selects $\sim 80$ percent of the
radio-excess sources defined using the total FIR.
Following \citet{2013A&A...549A..59D} we use $q_{100}<1.5$ to identify radio-excess sources in our
radio selected sample.
Clearly, such a criterion is simplistic and does not take into account the $z$ dependence, such that it may
suffer of contamination, especially at high redshift, by strong IR sources (i.e. hyper-luminous IR galaxies):
ideally, $q_{100}$ should be compared to the expected tracks for different representative 
types of galaxies. 
However, here we use the $q_{100}$ parameter not to build a clean sample of radio-excess sources, but rather to
mainly pinpoint embedded nuclear activity (RI-AGN) among the SF/hyb sources, identified via IR criteria.
Moreover, as we show below, the SF/hyb sources characterised by a radio excess are in a range of
redshift and  $q_{100}$ values supporting an AGN contribution to the observed radio emission 
regardless of the possible redshift evolution of the  $q_{100}$ parameter. 
A full IR SED
model fitting applied to a larger sample (selected from 1.4\,GHz observations) will be
presented in Paper II.

\begin{table*}
\caption{Summary of X-ray \& Radio-excess AGN}
\begin{center}
\begin{tabular}{lccccccc}
\hline
  IR Class       & \# X-ray & \# Radio-exc & \# X-ray \& Radio-exc.\\
\hline
 RE-AGN          & 10/19 & ~7/19 & 7/19 \\
 RE-AGN-candidate& ~8/13 & ~4/13 & 2/13 \\
 RI-AGN          & ~3/15 & 13/15 & 2/15 \\
 SF/hybrid       & ~9/30 & ~7/30 & 2/30 \\
\hline
\end{tabular}
\end{center}
\label{tab:xray}
\end{table*}

Fifty-two sources (out of the 77 radio sources with full IRAC photometry)
have a {\it Herschel-PACS} detection at 100\,$\mu$m, and for the remaining 25 objects
 we use a $3\sigma$ upper limit of 1.0\,mJy \citep{2013A&A...553A.132M, 2015A&A...573A..45M}.
The 1.4\,GHz flux densities are taken from
\citet{2010ApJS..188..178M} for all but three sources which are detected only at 5.5\,GHz.
For these three sources we derive the 1.4\,GHz flux density from that measured at 5.5\,GHz using a
spectral index of 0.7.

In Fig.\,\ref{plotq100} we plot the observed $q_{100}$ values against redshift
for all the catalogued sources with an infrared classification.
Considering $q_{100}<1.5$ as a proxy for selecting radio-excess sources, we have
31 radio-excess sources (41 percent of the sub-sample with full IRAC coverage).
As expected, radiatively-inefficient AGN
are typically associated with radio-excess sources: 87 percent (13/15) of this class of
sources have $q_{100}<1.5$ and for the two remaining sources the $q_{100}$ value
is an upper limit.
About one third (11/32) of RE-AGN (including also the RE-AGN-candidates),
are associated with radio-excess sources.
This significant fraction is not entirely surprising as
we are dealing with radio-selected RE-AGN.
However, the number of radio-excess RE-AGN
could vary significantly, since many of those at high redshift
are close to the $q_{100}$ threshold adopted here.
Moreover, at $z\gsim 2$ the criterion $q_{100}<1.5$ could be too simplistic
and a redshift evolution of this parameter should be taken into account.
In any case, this does not influence the census of nuclear activity as
most of the sources at such redshifts are already classified
as AGN on the basis of the IR colours or hard-band X-ray emission.
We also note that RE-AGN is the only class for which there is a clear
link between the radio-excess and the X-ray emission:
all the radio sources classified as RE-AGN and showing a
radio-excess (7 out of 19) are also X-ray AGN.

We find five radio sources at $z>1.5$, classified as SF/hyb systems
and not strong X-ray emitters, that
show a clear radio-excess ($q_{100}<1$ for four of them). 
Even taking into account the evolution with redshift of  $q_{100}$,
all these 5 sources would fall in the radio-excess region \citep[e.g. see Fig.5 in][]{2013A&A...549A..59D}.  
Finally, there is an IR-excess source at $z=2.5$: here 
the IR emission should be dominated by an AGN and indeed the sources is classified as such on the basis of both
the IR and X-ray diagnostics. It does not show a radio excess as typically observed in many RE-AGN.
The radio excess sources are shown as black squares in Fig.\,\ref{donley2}.

In Table\,\ref{tab:xray} we list, for each of the IR classes,
the number of X-ray AGN, the number of radio-excess sources and the number of sources which are both
X-ray and radio-excess AGN.
At this point it is important to recall that we have classified as RE-AGN-candidates those sources selected
as AGN by only one of the four IR diagnostics. The reason for this choice was to have
a separate class of sources potentially hosting nuclear activity for which there is a
significant possibility of contamination by non-AGN sources.
All but three (77 percent, 10/13) of the RE-AGN-candidates
\textit{are confirmed as AGN} by the X-ray luminosity or by their radio-excess.
We conclude that the difference between RE-AGN and RE-AGN-candidates is mainly due to the AGN dominance in the IR,
but it is reasonable to consider all these sources as true AGN. 

\subsection{1.4 GHz VLBI detections}
\label{vlbi}

The most direct confirmation of a radio AGN is
provided by the observation of high brightness temperature
components on milli-arcsec (mas) angular scales as probed by Very Long Baseline Interferometry (VLBI).
Although in the local Universe
compact and intense radio emission on mas-scales
might be also associated with Supernovae and Supernova Remnants,
these are difficult to detect with VLBI at $z>0.1$.

Twenty GOODS-N sources were detected by recent global VLBI observations at 1.6\,GHz
with S$_{VLBI} >50\,\mu$Jy and an angular resolution of 4\,mas \citep{2016A&A...587A..85R}.
Ten of them were previously
 detected with VLBI at 1.4\,GHz by \citet{2013A&A...550A..68C}.
VLBI cores are found to
account for, on average, 30 percent of the total 1.6\,GHz emission.
At 5.5\,GHz,  we catalogued 19 of the 20 sources
with a VLBI detection. The remaining source is located outside
the region covered by our 5.5\,GHz mosaic.
About half (9/19) of the 5.5\,GHz sources with VLBI detections
belong to the class of
RI-AGN and hence are associated with optically-passive galaxies.
This means that about 60 percent (9/15)
of the radio sources belonging
to this class have a VLBI detection, confirming the presence of radio bright compact cores in RI-AGN.
Regarding RE-AGN, about 26 percent (5/19) have a VLBI detection, with four of them
classified as radio-excess sources with a redshift $z\lsim 1$.
The remaining object is a RE-AGN with $q_{100}=1.7$ and  $z\sim 0.6$,
very close to the dividing line defining radio-excess sources).
Only one of the RE-AGN-candidates is detected by VLBI, and is also a radio-excess source and X-ray AGN.
Among the sources IR-classified as SF/hyb, three galaxies have a VLBI detection. All
three were classified as radio-excess sources and show a bright, compact core in the
VLA observations.
Finally, one source with a VLBI detection does not possess an IR classification.

\section{Discussion}
\label{disc}

In the previous section we investigated the AGN content of the radio catalogue, selected at 5.5\,GHz in the
GOODS-N field, via multi-wavelength AGN selection criteria.
The final classification scheme was obtained by combining all the criteria, as anticipated
in \S 6. Radio sources falling in at least one of the four IR AGN diagnostics
or which fulfil the X-ray luminosity requirement are
classified as RE-AGN.
Among the remaining sources, the RI-AGN are identified as those radio sources
having MIR colours typical of red and passive galaxies or those showing a radio-excess
(on the basis of the  $q_{100}$ parameter).
All the other sources,
which are not identified as AGN hosts by any of the criteria used here,
are classified as SFGs.

\begin{figure}
\centering
\includegraphics[width=8.5cm]{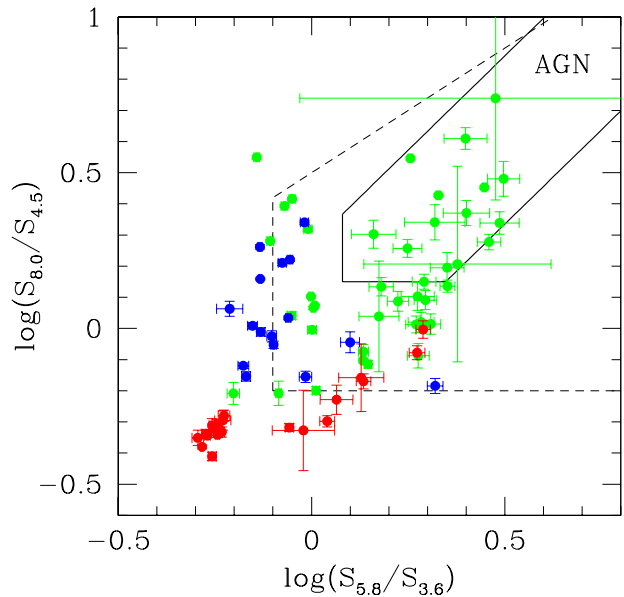}
\caption[]{ The IRAC colour-colour diagram for the 77
5.5\,GHz selected radio sources, detected at all IRAC bands.
The classification reported here has been
updated based on the X-ray luminosity and radio-excess criteria.
All the RE-AGN  (including the RE-AGN candidates) are shown with green symbols, the RI-AGN and SFGs
respectively with red and blue symbols.}
\label{donley_all}
\end{figure}

In Fig.\,\ref{donley_all} we plot the IRAC colour-colour diagram, already shown in Fig.\,\ref{midir1_plot},
updating the source classification and in Table\,\ref{tab_allclass} we list the 77 classified sources
indicating, together with the redshift,
the classification based on the four IR colour-colour plots, which source was classified as X-ray or
radio-excess AGN, or was VLBI detected, and the final classification as RE-AGN, RI-AGN or SFG.

\begin{table*}
\caption{Sample table listing the multi-wavelength classification for the first ten 5.5\,GHz sources with NIR identification.
The full version of the table is available as online-only material. Column\,1 gives the source name.
Column\,2 lists the spectroscopic ($^s$) or photometric ($^p$ ) redshift.
Column\,3 gives the classification based on the four IR colour-colour diagnostics.
Columns\,4 to 6: the crosses ("$\times$") identify the radio sources classified as X-ray AGN, radio-excess and VLBI sources.
Column\,7 is the final classification, determined by combining all the multi-wavelength diagnostics.
}
\label{tab_allclass}
\begin{center}
\begin{tabular}{lcccccc}
 \hline
  Source Name  &  $z$      &    class$_{\rm IR}$   &     X-ray     &     Radio-exc &      vlbi     &   class    \\
 \hline
 J123557+621536 & 0.433$^p$   &    SF/Hyb            &      -         &   -           &    -         &    SFG         \\
 J123601+621126 & 0.913$^s$   &    RI-AGN            &      -         &   $\times$    &    -         &    RI-AGN     \\
 J123603+621110 & 0.638$^s$   &    SF/Hyb            &      -         &   -           &    -         &    SFG         \\
 J123606+620951 & 0.772$^s$   &    SF/Hyb            &      $\times$  &   -           &    -         &    RE-AGN     \\
 J123606+621021 & 2.505$^s$   &    SF/Hyb            &      $\times$  &   $\times$    &    -         &    RE-AGN     \\
 J123608+621035 & 0.679$^s$   &    RE-AGN            &      $\times$  &   -           &    $\times$  &    RE-AGN     \\
 J123609+621422 & 0.779$^s$   &    SF/Hyb            &   -            &   -           &    -         &    SF          \\
 J123617+621011 & 0.846$^s$   &    SF/Hyb            &      $\times$  &   $\times$    &    -         &    RE-AGN     \\
 J123617+621540 & 1.993$^s$   &    SF/Hyb            &      -         &   $\times$    &    $\times$  &    RI-AGN     \\
 J123618+621550 & 2.186$^p$   &    RE-AGN$_{\rm cand}$ &      $\times$  &   $\times$    &    -         &    RE-AGN  \\
\hline
\hline
\end{tabular}
\end{center}
\end{table*}

Putting together all our multi-wavelength AGN classifications,
the first notable thing is the large fraction of AGN in the 5.5\,GHz catalogue:
about 79 percent (61/77) of the IR-classified sources
show evidence for nuclear activity, in at least one of the radio, infrared or X-ray bands.
This fraction becomes even higher, 95 percent (35/37),
when considering only the radio sources with $z>1.5$.
This fraction of AGN is very large,
especially if we consider that we are sampling a
population of faint radio sources: the median peak brightness and total flux densities
for the 77 sources with full IRAC coverage are about 20\,$\mu$Jy beam$^{-1}$ and 40\,$\mu$Jy at 
5.5\,GHz, respectively.
Moreover, we  note that radio sources classified as AGN (both RE and RI-AGN)
dominate at all flux density levels.
Other radio surveys with comparable sensitivity, e.g. E-CDFS \citep{2013ApJS..205...13M, 2013MNRAS.436.3759B}
or VLA-COSMOS 3GHz \citep{2017arXiv170309713S, 2017arXiv170309720D},
derive a fraction of radio detected AGN of about 40 percent, a factor of two lower than that derived in this paper. 

As argued below, we think that the large fraction of AGN is a selection effect mainly due to the
the lack of (or limited) short spacing information in our
VLA (A-array dominated) data that limits
the largest scale structure that can be imaged, introducing a bias against extended 
($>1-2$\,arcsec) low-surface brightness sources.
Indeed, AGN- and star-formation-related radio emission should display distinct morphological structures.
In particular, radio sources hosting nuclear activity should preferentially have a compact component,
while star-forming galaxies should be characterised by extended/diffuse radio emission on kpc scales,
associated with the galactic disk.
We tested this by analysing the deconvolved angular size
of the sources derived from the source fitting procedure (see \S\,\ref{catalogue}).
We used the major axis as an estimate of the source size.
For those sources that are classified
as unresolved, on the basis of the relation between the total-to-peak ratio and the SNR,
we assumed that the fitted size is an upper limit.
For a more homogeneous comparison, we restricted this analysis to the sources with redshift $z<1.5$,
since 14 of the 16 SFGs are within this limit.
We derived the Kaplan-Meier median estimator using the {\sc ASURV} package.
We find that RE-AGN and RI-AGN have the same median sizes, and for this reason we  combine all the
AGN, deriving a median size of $0.29\pm 0.22$ arcsec for the AGN and
$0.79\pm 0.21$ arcsec for the SFGs (the quoted error is the MAD).
This result is consistent with our classification based on  the IR colours, X-ray luminosities or
radio-excess. Indeed, the median angular size
of the SFGs corresponds, at $z=1$, to $\sim 6$ kpc, consistent with a radio emission distributed over a galactic
disk.

So far we have found that, among the sources detected at 5.5\,GHz, those classified as AGN
are  more compact than those classified as star-forming galaxies. 
On the other hand, the overwhelming fraction of AGN in our 5.5\,GHz-selected sample are,
apparently, at odds with the results of other deep radio surveys and the classifications reported in
\citet[][hereafter M05]{2005MNRAS.358.1159M} for a complete sample selected at 1.4\,GHz in GOODS-N.
The sample in M05 contains 92 sources with flux densities at 1.4\,GHz above
40\,$\mu$Jy, from a 10$\times$10\,arcmin$^2$  region,
within the area covered by our 5.5\,GHz VLA observations.
The classification adopted in M05 divided the radio sources in secure or candidate AGN or SFGs
on the basis of the radio/optical combined morphology, radio spectral index, X-ray luminosity, and ISO detection.
We emphasize that the criteria adopted in M05 to classify a radio source as AGN or SFG
are either different to those applied in this paper or based on shallower data at X-ray and infrared wavelengths.
Adding together the candidates and the secure classifications, both for AGN and SFGs,
more than half of the sources in M05 are classified as SFGs
(48/92, $52$ percent), while only one fifth are classified as AGN (18/92, $20$ percent). 
The remaining sources (26/92, $28$ percent) are unclassified, meaning that
the radio properties could be  associated either with AGN or starburst activity.
In principle, by assuming a radio spectral index of 0.7, we should be able to detect
89 out of the 92 sources ($\simeq 97$ percent) listed in M05,
at the point source sensitivity of our 5.5\,GHz mosaic.
In practice,  since most of the sources are resolved at the angular resolution of our 5.5\,GHz image,
we detect only $\sim$ 60 percent of the 1.4\,GHz selected complete sample.

Using the values listed in Table A2 from M05 we derived the median values of the 1.4\,GHz total flux and
largest angular size for the sources detected and not detected at 5.5\, GHz,
Out of the 48 SFGs classified by M05, 25 are detected at 5.5\,GHz.
These are the objects
with higher flux density and smaller sizes: for the 25 sources detected at 5.5\,GHz the
median 1.4\,GHz flux density and median largest angular size
(LAS)  are 71\,$\mu$Jy and 0.8\,arcsec, compared with
53 $\mu$Jy and 1.2 arcsec for the 23 radio sources undetected at 5.5\,GHz.
The same is observed for the sources unclassified by M05: 14 out of 26 are detected at
5.5\,GHz with a median flux density and LAS of 124\,$\mu$Jy and 0.8\,arcsec,
compared to 72\,$\mu$Jy and 2.1\,arcsec for those undetected at 5.5\,GHz. 
On the other hand,  16 out of 18 AGN classified by M05 are detected at 5.5. These 16 sources have a median 
1.4\,GHz flux density and LAS of 217\,$\mu$Jy and 0.6\,arcsec. The only two AGN undetected at 5.5\,GHz are
the two weakest AGN with LAS $\ge 2.5$ arcsec in M05. 

We conclude that we are systematically missing faint sources with sizes $\gsim 1$ arcsec, and that these sources
are usually identified with SFGs in M05. This is not surprising, since while the point source sensitivity of our 
observations at 5.5\,GHz and that used by M05 at 1.4 are comparable (once scaled to take into account the radio 
spectral  index) the two surveys have different beam solid angles ($0.56\times 0.47$ arcsec$^2$ at 5.5\,GHz and 
$2\times 2$ arcsec$^2$ at 1.4\,GHz), and 
therefore different brightness sensitivities: the lower resolution observations in M05 are about 15 times as sensitive
as our high resolution observations for sources with sizes larger than the beam \citep[e.g.][]{2015arXiv150205616C}.
Even the VLA-COSMOS 3GHz survey, that is the closest in frequency and resolution ($0.75\times 0.75$ arcsec) to our
observations is five times as sensitive as our observations in terms of brightness sensitivity. 
If we assume that the 35 SFGs and unclassified sources in M05, not detected at 5.5\,GHz, are indeed all 
SFGs, the overall fraction of AGN in our sample becomes less extreme and more in line with
expectation at the tens of $\mu$Jy level.
As pointed out above, short spacing data are required to properly sample low-surface brightness sources
on arcsec scale: this would be achievable by adding VLA C-configuration
observations to our data. Nonetheless, our results  highlight the usefulness of
the present 5.5\,GHz observations in selecting radio-emitting AGN at the faintest flux levels
($\lsim 100\,\mu$Jy).

One further point that needs to be mentioned is the different classifications in M05 and the present paper.
We note that while all the sources classified as AGN in M05 and detected at 5.5\,GHz are also classified as
AGN in the present paper, this is not the true for sources previously classified as SFGs. The majority of
the sources classified as SFGs or unclassified by M05 and detected in our observations turn out to be classified
as AGN by our criteria. 
This can be explained by the deeper infrared and X-ray observations used in the present paper
that allowed to detect lower luminosity AGN with respect to the ancillary data used in M05.

\section{Conclusions}

Using ultra-deep sub-arcsec-resolution radio observations at 5.5\,GHz obtained
with the VLA in the framework of the {\it e}MERGE legacy project, we have produced a
mosaic (including seven different pointings) with a median r.m.s. noise of
3\,$\mu$Jy\,beam$^{-1}$ and an angular resolution of $0.56\times 0.47$ arcsec$^2$
over a circular region with a diameter of 14\,arcmin centred on the
GOODS-N radio field.

The main results presented in this paper can can be summarised
as follows:

\begin{itemize}

\item We extracted a catalogue containing 94 radio sources above the local $5\sigma$
threshold, with about 50 percent of the sources in the range $10-30$\,$\mu$Jy beam$^{-1}$
and less than 20 percent with peak flux $> 100$\,$\mu$Jy\,beam$^{-1}$. About
60 percent (56/94) of the radio sources are classified as resolved on the basis
of the total-to-peak flux ratio versus SNR plot.

\item
We used deep NIR catalogues, mainly \citet{2010ApJS..187..251W}, but also \citet{2011PASJ...63S.379K}
and \citet{2014ApJS..214...24S},
to identify the radio sources. We find that 88 percent (83/94)
of the radio catalogue have secure NIR identifications, with the
fraction raising to 96 percent (76/79) when only the radio sources above
$5.5\sigma$ are considered.

\item
Redshift information is available for 95 percent (79/83) of the NIR identified
radio sources (51 redshifts are spectroscopic and 28 photometric).
The median redshift is $z_{med}=1.32$.

\item
We used multi-band AGN diagnostics (IR colour-colour plots, X-ray luminosity,
radio-excess parameter and VLBI detection) to separate AGN-driven
(both radiatively efficient and inefficient) radio sources from
SFGs in a subsample of 77 radio sources with a detection in all the four
IRAC bands. We find that 79 percent (61/77) of the sources show evidence for nuclear activity
and this fraction is about 92 percent if we consider only the sources with redshift $z>1.5$.
Such a large fraction of AGN is unusual considering we are sampling a population of radio sources
with a median peak brightness of $\simeq$ 20\,$\mu$Jy\,beam$^{-1}$.

\item
Our conclusion is that we are missing SFGs because of
the limited surface brightness sensitivity, due to the limited availability of short
spacings. This favours the detection of compact kpc/sub-kpc radio sources at the expense of sources
with radio emission  distributed on scales of several kpc.
Indeed, the AGN populations (both RE- and RI-AGN) have very similar
median angular sizes ($\simeq 0.2-0.3$\,arcsec for $z<1.5$), the SFGs have larger sizes
($\simeq 0.8$ arcsec for $z<1.5$).
Finally, AGN-hosting radio sources  (RE- and RI-AGN) dominate the population of our catalogue
at all flux density levels.

\end{itemize}

The afore-mentioned selection effects need to be taken into account in planning future surveys.
Such effects
will be further discussed in a forthcoming paper based on
a comparative analysis of radio-selected samples with different angular resolutions and frequency,
but comparable depths to GOODS-N.
In that paper we will also discuss the origin of the radio emission in RE-AGN.

\section*{Acknowledgements}
DG, MB, IP acknowledge support from PRIN-INAF 2014 (PI M. Bondi).
DG and IP acknowledge support of the Ministry of Foreign Affairs and
International Cooperation, Directorate General for the Country Promotion
(Bilateral Grant Agreement ZA14GR02 - Mapping the Universe on the Pathway
to SKA)


\end{document}